\documentstyle[12pt]{article}

\def\presentation{
\voffset -.70in \hoffset -.19in
\oddsidemargin 0in \evensidemargin 0in
\marginparwidth .75in \marginparsep 7pt \topmargin 0in
\headheight 12pt \headsep .25in
\footheight 18pt \footskip .35in
\textheight 9.5in \textwidth 6.5in
\columnsep 10pt \columnseprule 0pt }

\presentation
\begin{document}

%
\def\tilde{\widetilde}
\def\bar{\overline}
\def\hat{\widehat}
\def\*{\star}
\def\({\left(}		\def\BL{\Bigr(}
\def\){\right)}		\def\BR{\Bigr)}
\def\[{\left[}		\def\BBL{\Bigr[}
\def\]{\right]}		\def\BBR{\Bigr]}
\def\lb{\[}
\def\rb{\]}
%
%
\def\frac#1#2{{#1 \over #2}}		\def\x{ \otimes }
\def\inv#1{{1 \over #1}}
\def\half{{1 \over 2}}
\def\d{\partial}
\def\der#1{{\partial \over \partial #1}}
\def\dd#1#2{{\partial #1 \over \partial #2}}
\def\vev#1{\langle #1 \rangle}
\def\ket#1{ | #1 \rangle}
\def\bra#1{ \langle #1 |}
\def\rvac{\hbox{$\vert 0\rangle$}}
\def\lvac{\hbox{$\langle 0 \vert $}}
\def\comm#1#2{ \BBL\ #1\ ,\ #2 \BBR }
\def\2pi{\hbox{$2\pi i$}}
\def\e#1{{\rm e}^{^{\textstyle #1}}}
\def\grad#1{\,\nabla\!_{{#1}}\,}
\def\dsl{\raise.15ex\hbox{/}\kern-.57em\partial}
\def\Dsl{\,\raise.15ex\hbox{/}\mkern-.13.5mu D}
%
%
\def\th{\theta}		\def\Th{\Theta}
\def\ga{\gamma}		\def\Ga{\Gamma}
\def\be{\beta}
\def\al{\alpha}
\def\ep{\epsilon}
\def\la{\lambda}	\def\La{\Lambda}
\def\de{\delta}		\def\De{\Delta}
\def\om{\omega}		\def\Om{\Omega}
\def\sig{\sigma}	\def\Sig{\Sigma}
\def\vphi{\varphi}
%
%
\def\CA{{\cal A}}	\def\CB{{\cal B}}	\def\CC{{\cal C}}
\def\CD{{\cal D}}	\def\CE{{\cal E}}	\def\CF{{\cal F}}
\def\CG{{\cal G}}	\def\CH{{\cal H}}	\def\CI{{\cal J}}
\def\CJ{{\cal J}}	\def\CK{{\cal K}}	\def\CL{{\cal L}}
\def\CM{{\cal M}}	\def\CN{{\cal N}}	\def\CO{{\cal O}}
\def\CP{{\cal P}}	\def\CQ{{\cal Q}}	\def\CR{{\cal R}}
\def\CS{{\cal S}}	\def\CT{{\cal T}}	\def\CU{{\cal U}}
\def\CV{{\cal V}}	\def\CW{{\cal W}}	\def\CX{{\cal X}}
\def\CY{{\cal Y}}	\def\CZ{{\cal Z}}
%
%
\font\numbers=cmss12
\font\upright=cmu10 scaled\magstep1
\def\stroke{\vrule height8pt width0.4pt depth-0.1pt}
\def\topfleck{\vrule height8pt width0.5pt depth-5.9pt}
\def\botfleck{\vrule height2pt width0.5pt depth0.1pt}
\def\Zmath{\vcenter{\hbox{\numbers\rlap{\rlap{Z}\kern 0.8pt\topfleck}
		\kern 2.2pt \rlap Z\kern 6pt\botfleck\kern 1pt}}}
\def\Qmath{\vcenter{\hbox{\upright\rlap{\rlap{Q}\kern
                   3.8pt\stroke}\phantom{Q}}}}
\def\Nmath{\vcenter{\hbox{\upright\rlap{I}\kern 1.7pt N}}}
\def\Cmath{\vcenter{\hbox{\upright\rlap{\rlap{C}\kern
                   3.8pt\stroke}\phantom{C}}}}
\def\Rmath{\vcenter{\hbox{\upright\rlap{I}\kern 1.7pt R}}}
\def\Z{\ifmmode\Zmath\else$\Zmath$\fi}
\def\Q{\ifmmode\Qmath\else$\Qmath$\fi}
\def\N{\ifmmode\Nmath\else$\Nmath$\fi}
\def\C{\ifmmode\Cmath\else$\Cmath$\fi}
\def\R{\ifmmode\Rmath\else$\Rmath$\fi}

\rightline{SPhT-91-166; LPTHE-91-56}
\vskip 2cm
\centerline{\LARGE Dressing Symmetries.}
\vskip1cm
\centerline{\large Olivier Babelon }
\centerline{Laboratoire de Physique Th\'eorique et Hautes
Energies \footnote[1]{\it Laboratoire associ\'e au CNRS.}}
\centerline{
 Universit\'e Pierre et Marie Curie, Tour 16 1$^{er}$
\'etage, 4
place Jussieu}
\centerline{75252 Paris cedex 05-France}
\vskip1cm
 \centerline{\large  Denis Bernard }
 \centerline{Service de Physique Th\'eorique de Saclay
\footnote[2]{\it Laboratoire de la Direction des Sciences de la
Mati\`ere du Commisariat \`a l'Energie Atomique.}}
\centerline{F-91191, Gif-sur-Yvette, France.}
 \vskip3cm
Abstract.\\
We study Lie-Poisson actions on symplectic manifolds. We
show that they are generated by non-Abelian Hamiltonians.
We apply this result to
the group of dressing transformations in soliton theories;
we find that the non-Abelian Hamiltonian
is just the monodromy matrix. This provides a new proof of their Lie-Poisson
property. We show that the dressing transformations are the classical
precursors of the non-local and quantum group symmetries of these theories.
We treat in detail the examples of the Toda field
theories and the Heisenberg model.

 \vfill
  \newpage

\def\x{\stackrel{\otimes}{,}}
\def\proof{\noindent Proof. \hfill \break}
\def\Gr{ G_R }
\def\gr{ \CG_R }
\def\alv{ \alpha^\vee }
\def\xib{ {\bar \xi} }
\def\oti{ \otimes }
\def\otp{ {\oti_,} }
\def\lam{ \la_{{\rm max}} }
\def\debut{ \begin{eqnarray} }
\def\fin{ \end{eqnarray} }
\def\non { \nonumber\\ }

\noindent {\LARGE  0 ~~ Introduction.}

\bigskip
\noindent Are quantum integrable models reducible to quantum group theory?
This assertion is best exemplified by the algebraic formulation
of two-dimensional conformal field theories \cite{BPZ} but remains uncertain
for massive integrable models. On one hand, quantum groups \cite{Dr86,Ji85}
and their representation theory are now well understood although
some of their relations with the quantum algebraic Bethe Ansatz
are still mysterious. On the other hand, a large number of methods
for studying  integrable models have been developed,
cf e.g. \cite{FaTa86,Ga0}, including
the most famous algebraic quantum inverse scattering method \cite{Fa82}.
More recently, new developments have suggested the use of non-local
symmetries for reformulating integrable models as quantum group theories
\cite{B91,BL91,S91}. However, despite suggestive facts, the assertion is
still not established by any of these approaches.

The aim of this paper is to try to understand the occurrence of non-local
symmetries in classical soliton equations and their relations with
semi-classical analogues of quantum group symmetries.

We will deal with integrable soliton equations having a zero-curvature
representation and admitting a Hamiltonian formulation based on the
standard transfer matrix formalism. As is well known, this formulation
ensures that the models possess integrals of motion in involution.
These integrals are computed from the monodromy matrix $T$ as:
\debut
H_{involution} = tr(T) \nonumber
\fin
Natural candidates for the non-local
symmetries are the so-called dressing transformations
\cite{ZaSh79,DaJiKaMi81,Se85}. These are special
 transformations acting on the solution space of the soliton equations
or equivalently on the phase space. They  form a usually infinite dimensional
symmetry group of the soliton equations. Moreover, this group possesses
three remarkable properties:

\noindent {\bf (a) Dressings are Lie-Poisson actions.}
In Hamiltonian mechanics one is usually used to consider symplectic
actions generated by Hamiltonians. Lie-Poisson actions are
generalizations of this notion. Namely, they are actions of
a Lie group, itself equipped with a Poisson structure, such that the
Poisson brackets transform covariantly.
Contrary to the symplectic actions, the Lie-Poisson actions
are not generated by Hamiltonian functions but, as we will show,
by non-Abelian generalizations of them, which we refer to as
non-Abelian Hamiltonians. This non-Abelian structure
reflects the nature of the Poisson bracket on the Lie group which,
in its turn, is the semi-classical ancestor of the quantum group.
The remarkable property of the group of dressing transformations
is that, if it is equipped with suitable
Poisson brackets, it defines a Lie-Poisson action
on the phase space \cite{Se85}.
Moreover, as we will prove, the non-Abelian Hamiltonian of these
transformations is the monodromy matrix of the models.
This means that the generators of the infinitesimal
dressing transformations can be thought of as being:
\debut
H_{dress} = tr( X \log T)\nonumber
\fin
for any $X$ in the Lie algebra of the dressing group. (A more precise
formulation will be given in the following.)
This relation indicates
that the transfer matrix can be
reconstructed from the ``Hamiltonians" $H_{dress}$ and
$H_{involution}$.

\noindent {\bf (b) Dressings are non-local.}
This is true by construction.
It implies that the charges which generate the dressing transformations
are also non-local. This non-locality is an echo of the Poisson
structure defined on the dressing group. Both facts have two
related consequences: (i) the dressing transformations are not linearly
generated by the charges, and (ii) the Poisson algebra of the charges
is not the Lie algebra of the dressing group although it is related:
It is a semi-classical deformation of it.
However, any algebraic relation in the dressing Lie algebra, e.g. the
Serre-like relations, yields a relation in the dressing Poisson algebra.

\noindent {\bf (c) Dressings induce field multiplets.}
Because the dressing transformations map solutions of the soliton
equations into solutions, they provide a way to gather the fields
into non-local field multiplets \cite{BF91,LS91}, alternatively named
non-local blocks. For example, by dressing local
conserved currents one defines a set of non-local conserved currents
which form the current multiplets generating the dressing transformations.
The field multiplets are closed by dressing and form orbits of the
dressing group. The number of multiplets is therefore the number
of orbits of the dressing group in the phase space. We do  not
know under which conditions the quotient
of the phase space by the dressing group reduces to a finite
number of points. (This notion is the classical analogue
of the notion of rationality in  massive quantum field theory.)
However, a few hints indicate that this is the
case in some models invariant under infinite dressing groups.

Section 1 deals with basic facts about Lie-Poisson actions. We
generalize the notion of the moment map and explain its relation
with the above mentioned non-Abelian Hamiltonians. We give new
proofs of the Lie-Poisson properties of the dressing transformations
and establish the relation between the non-Abelian Hamiltonians
of these transformations and the monodromy matrices

Section 2 is concerned with the dressing transformations in the Toda
field theories. In particular, we show how they provide a classical
explanation of the occurence of quantum group symmetries in
two-dimensional conformal field theories.
In this case the orbits of the dressing
group are in one-to-one correspondence with the
fundamental highest weights of the Lie algebra.

In Section 3 we study the Heisenberg model. We show that the
Poisson algebra of the dressing transformations is a semi-classical
Yangian. We describe the relation between the local conserved quantities
and the generators of dressing transformations.
In this case the orbit space is likely to
be reduced to a finite number of points (as local arguments indicate).
Moreover, the relation between the algebraic Bethe Ansatz and
the representation theory of the dressing group appears clearly.

Although dressing transformations and $\cal T$au functions are
intimately related, e.g. $\cal T$au functions are orbits
of the dressing groups, we reserve their study for a future work.

\section{Lie Poisson Actions and Dressing Transformations.}

In this section we present a few basic facts about Lie-Poisson
actions and dressing transformations. Some of them
also appeared in \cite{Dr86,Se85,Lu90,FaGa91}.
However, we feel that it is worth
explaining these results in a way perhaps more accessible
to a physicist. We will also give new proofs of these results.
Let $M$ be a sympletic manifold. We denote by $\{\ ,\ \}_M$
the Poisson bracket in $M$.

\bigskip
\noindent {\bf 1a- Sympletic actions.}
Before describing Lie-Poisson actions,
we recall some well known facts about Hamiltonian actions
\cite{Ar76,AbMa78}.  Let $H$ be a Lie group and
$\CH$ its Lie algebra. The action of
a one parameter subgroup $(h^t)$ of $H$ is said to be symplectic if for any
functions $f_1$ and $f_2$ on $M$ ,
\begin{eqnarray}
\{ f_1(h^t.x), f_2(h^t.x)\}_M\ =\ \{f_1,f_2\}_M(h^t.x) \label{EIi}
\end{eqnarray}
Introducing the vector field $X$ on $M$ corresponding to the infinitesimal
action, $X.f(x) = \frac{d}{dt}f(h^t.x)|_{t=0}$, the condition (\ref{EIi})
becomes:
\begin{eqnarray}
\{ X.f_1 , f_2 \}_M\ +\
\{ f_1 , X.f_2 \}_M\ =\ X.\{f_1,f_2\}_M \label{EIii}
\end{eqnarray}
We have the standard property that the action of any one parameter
subgroup of $H$ is locally Hamiltonian. This means that there
exists a function $H_X$, locally defined on $M$, such that:
\begin{eqnarray}
X.f\ =\ \{H_X, f\}_M \label{EIiii}
\end{eqnarray}
The proof is standard and is a particular case of the more general
proof we will describe in the following section. The global existence
of $H_X$ is another state of affair.

The Hamiltonians $H_X$ are used to define the moment map. Namely,
assume that they are globally defined on $M$, depend
linearly on $X\in\CH$, and are such that:
\begin{eqnarray}
H_{[X,Y]} = \{H_X , H_Y\}_M \nonumber
\end{eqnarray}
Then, the moment map $P$ is a map from $M$ to $\CH^*$, defined by:
\begin{eqnarray}
P: M &\longrightarrow & \CH^* \nonumber \\
 P(x)(X)&=&H_X(x)\qquad;\qquad \forall\ X\in\CH \nonumber
\end{eqnarray}
Moreover, the symplectic action of $H$ on $M$ is transformed by
$P$ into the coadjoint action of $H$ on $\CH^*$:
\begin{eqnarray}
H_X(h.x)\ =\ H_{(Ad_Hh).X}(x),\qquad
\forall\ x\in M\ ;\ \forall\ X\in\CH
\ ;\ \forall\ h\in H \nonumber
\end{eqnarray}
These are the properties we want to generalize to Lie-Poisson actions.

\bigskip
\noindent {\bf 1b- Poisson-Lie groups.} Assume now that $H$ is a
Poisson- Lie group. This means that $H$ is a Lie group equipped with a
Poisson structure such that the multiplication in $H$ viewed as map
$H\times H \to H$ is a Poisson mapping.
Let us be more explicit. Any Poisson bracket $\{,\}_H$ on a Lie
group $H$ is uniquely characterized by the data of a
$\CH\oti\CH$-valued function: $h\in H\to \eta(h)\in\CH\oti\CH$.
Indeed, introducing a basis $(e_a)$ of $\CH$, the Poisson
bracket  for any functions
$f_1$ and $f_2$ on $H$ can be written as :
\begin{eqnarray}
\{f_1,f_2\}_H(h)=\sum_{a,b}\ \eta^{ab}(h)(\nabla^R_af_1)(h)
		(\nabla^R_bf_2)(h) \label{Epi}
\end{eqnarray}
where $\eta(h)=\sum_{a,b}\eta^{ab}(h)e_a\oti e_b$ and
$\nabla^R_a$ is the right-invariant vector field corresponding
to the element $e_a\in\CH$:
\begin{eqnarray}
\nabla^R_a f(h)= { {d\over{dt}}f(e^{te_a}h)}\Big\vert_{t=0} \nonumber
\end{eqnarray}
In particular, if we take for $f$ the matrix elements of a
representation of $H$, we get
\begin{eqnarray}
\{h\x h\}_H = \eta(h)\cdot\ h\oti h
\qquad;\qquad h\in H \label{Epii}
\end{eqnarray}
The antisymmetry of the Poisson bracket (\ref{Epi})
requires $\eta_{12}=-\eta_{21}$, and the Jacobi identity is equivalent
to a quadratic relation for $\eta$ which can be easily written down.
The Lie Poisson property of the Poisson brackets (\ref{Epii}) is the
requirement that they transform covariantly under the multiplication
in $H$; e.g. for the right multiplication $\rho^g(h)=hg$
this corresponds to the equality of the following brackets:
\debut
\{\rho^g(h)\x \rho^g(h)\}_{H\times H} &=& \{h\x h\}_H\ g\oti g\
+\ h\x  h\ \{g\x g\}_H \nonumber\\
\rho^g(\{h\x h\}_H) &=& \eta(hg)\cdot hg\oti hg \nonumber
\fin
Comparing these two equalities implies that $\eta(h)$ is a cocycle \cite{Dr86}:
\debut
\eta(hg) = \eta(h) + Ad\, h\cdot \eta(g) \label{Epiv}
\fin
The same cocycle condition is needed for the left multiplication
to be a Lie Poisson mapping.

The bracket $\{,\}_H$ can be used to define a Lie algebra
structure on $\CH^*$ by the following formula:
\begin{eqnarray}
\[ d_e\phi_1, d_e\phi_2 \]_{\CH^*}\ =\ d_e\{\phi_1,\phi_2\}_H
\label{EIvii}
\end{eqnarray}
with $d_e\phi\ \in\CH^*$ the differential of the function $\phi$
on $H$ evaluated at the identity of $H$.
Introducing a basis $(e^a)$ in $\CH^*$, dual to
the basis $(e_a)$ in $\CH$, the differential at the identity can
written as $d_e\phi=\sum_a e^a(\nabla^L_a\phi)(e)\in\CH^*$ where
$\nabla^L_a$ are the left-invariant vector fields on $H$ i.e
\debut
\nabla^L_a f(h)={d\over{dt}} f(he^{te_a})\Big\vert_{t=0}\nonumber
\fin
In this basis, eq.(\ref{EIvii}) gives:
\debut
\[e^a,e^b\]_{\CH^*}\ =\ f^{ab}_c\ e^c \label{Epv}
\fin
where the structure constants are $f^{ab}_c=(\nabla^L_c\eta^{ab})(e)$.
The Lie bracket eq.(\ref{Epv}) satisfies the Jacobi identity thanks to the
Jacobi identity for the Poisson bracket in $H$.
We denote by $H^*$ the Lie group with Lie algebra $\CH^*$.

In the same way as the Poisson structure on $H$ induces a Lie algebra
structure on $\CH^*$, the Lie algebra structure on $\CH$ induces
a Lie-Poisson structure on $H^*$. We denote by $\{,\}_{H^*}$ these
Poisson brackets and by $\eta^*$ the corresponding cocycle:
$\eta^*(\ga)=\sum_{a,b}\eta^*_{ab}(\ga)e^a\oti e^b
\in\CH^*\oti\CH^*$. It is characterized by the property that:
\debut
(\nabla_L^d\eta^*_{ab})(e)\ =\ C^d_{ab}\nonumber
\fin
where $C^d_{ab}$ are the
structure constants of $\CH$ in the basis $(e_a)\in\CH$.

\bigskip
\noindent {\bf 1c- The action of a Poisson-Lie
group on a symplectic manifold.}
The action of a Poisson-Lie group on a symplectic manilfold is a Lie-Poisson
action if the Poisson brackets transform covariantly;
i.e. if for any $h\in H$ and any function $f_1$ and $f_2$ on $M$,
\begin{eqnarray}
\{f_1(h.x),f_2(h.x)\}_{H\times M}\ =\
\{f_1,f_2\}_M(h.x) \label{EIviii}
\end{eqnarray}
The Poisson structure on $H\times M$ is the product Poisson structure.

Let $X\in\CH$ and denote also by $X$ the vector field on $M$ corresponding
to the infinitesimal transformation generated by $X$. By duality, this defines,
for any function $f$ on M, a covector $\zeta_f$, taking values in $\CH^*$
by:
\begin{eqnarray}
X.f(x)\ =\ \frac{d}{dt}f(e^{tX}.x)|_{t=0}\ =\
<\zeta_f(x),X> \nonumber
\end{eqnarray}
where $<,>$ denote the pairing between $\CH$ and $\CH^*$.
The infinitesimal form of (\ref{EIviii}) is then:
\begin{eqnarray}
\{ X.f_1 , f_2 \}_M\ +\
\{ f_1 , X.f_2 \}_M\ +\ <[\zeta_{f_1},\zeta_{f_2}]_{\CH^*},X>\
=\ X.\{f_1,f_2\}_M \label{EIx}
\end{eqnarray}
or equivalently,
\begin{eqnarray}
\{ \zeta_{f_1} , f_2 \}_M\ +\
\{ f_1 , \zeta_{f_2} \}_M\ +\ \[{\zeta_{f_1},\zeta_{f_2}}\]_{\CH^*}\
=\ \zeta_{\{f_1,f_2\}_M} \nonumber
\end{eqnarray}

Introducing two dual basis of the Lie algebras $\CH$ and $\CH^*$,
$e_a\in\CH$ and $e^a\in\CH^*$ with $<e^a,e_b>=\delta^a_b$,
eq. (\ref{EIx}) becomes:
\begin{eqnarray}
\{e_a.f_1,f_2\}_M\ +\
\{f_1,e_a.f_2\}_M\ +\ f_a^{bd} (e_b.f_1)(e_d.f_2)\ =\
e_a.\{f_1,f_2\}_M \label{EIxi}
\end{eqnarray}
It follows immediately from eq.(\ref{EIx}) that a Lie-Poisson action cannot
be Hamiltonian unless the algebra $\CH^*$ is Abelian.\\
 However, in general, we have a non-Abelian analogue of
the Hamiltonian action eq~.~(\ref{EIiii})~\cite{Lu90}.
\proclaim  Proposition.
There exists a function
$\Ga$, locally defined on $M$ and taking values in the group $H^*$,
such that for any function $f$ on $M$,
\begin{eqnarray}
X.f\ =\ <\ \Ga^{-1}\ \{f,\Ga\}_M,X>\quad,\quad \forall\ X\in\CH
\label{EIxii}
\end{eqnarray}
or equivalently,
\debut
\zeta_f(x)\ =\ \Ga^{-1}\ \{f,\Ga\}_M(x)\nonumber
\fin
We will refer to $\Ga$ as the non-Abelian Hamiltonian of the Lie-Poisson
action.\par
\noindent The proof is the following. Introduce the Darboux coordinates
$(q^i,p^i)$.
Let $\Om= e^a\ \Om_a$ be the $\CH^*$-valued one-form
defined by $\Om_a=e_a^{q^i}dp^i\ -\ e_a^{p^i}dq^i$ where
$e_a^{q^i}$, $e^{p^i}_a$ are the components of the vector field $e_a$,
$e_a=e_a^{q^i}\d_{q^i}+e_a^{p^i}\d_{p^i}$. Eq. (\ref{EIx}) is then
equivalent to the following zero-curvature condition for $\Om$:
\begin{eqnarray}
d\Om\ + \[\Om,\Om\]_{\CH^*}\ =\ 0 \nonumber
\end{eqnarray}
Therefore, locally on $M$, $\Om=\Ga^{-1}\ d\Ga$.
This proves eq.(\ref{EIxii}).\\
The converse is true: an action generated by a non-Abelian
Hamiltonian as in eq.(\ref{EIxii}) is Lie-Poisson since then we have:
\begin{eqnarray}
X.\{ f_1 ,f_2 \}_M &-& \{X.f_1 , f_2 \}_M - \{f_1 , X.f_2 \}_M \nonumber \\
&=& <\BBL \Gamma^{-1} \{ f_1 ,\Ga \}_M ~,~
\Ga^{-1}\{f_2 ,\Ga \}_M \BBR _{\CH^*} ,X > \nonumber
\end{eqnarray}

The moment map $\CP$ for the Lie-Poisson action is defined by:
\begin{eqnarray}
\CP: M &\longrightarrow & H^*  \label{EIxiv} \\
  x &\longrightarrow & \Ga(x)\
\qquad {\rm with}\qquad \Ga(x)\ =\ \exp\({-Q(x)}\).\nonumber
\end{eqnarray}
We refer to the  $\CH^*$-valued functions $Q(x)$ as the charges generating
the Lie-Poisson action.

The transformation (\ref{EIxii}) can
alternatively be written in terms of the charges $Q$ as follows.
Let $\om^a$ be the component of the Maurer-Cartan form in $H^*$,
$\ga^{-1}d\ga= \om^a\ dq_a$, in the local coordinate system $q_a$
specified by the exponential map, $\ga=\exp(-q_ae^a)\ \in H^*$. Let
$Q_a$ and $\Om^a$ be the pull-back of $q_a$ and of $\om^a$ by $\CP$,
i.e. $Q_a=\CP^*q_a$, $\Ga=\exp(-Q_ae^a)=\CP^*\ga$ and $\Om^a=\CP^*\om^a$,
 then,
\begin{eqnarray}
X.f(x)\ =\ <\Om^a(Q(x))\ \{Q_a,f\}_M (x),X> \label{EIxv}
\end{eqnarray}
Or explicitely,
\begin{eqnarray}
X.f(x)\ =\
<\[{\frac{\exp(Ad_{\CH^*}Q(x))-1}{Ad_{\CH^*}Q(x)}}\]\cdot
\ \bigl\{Q,f\bigr\}_M (x)\ ,X> \nonumber
\end{eqnarray}

As $\Ga$ is the pull-back of an element $\ga\in H^*$ by $\CP$,
it is natural to require for the Poisson brackets in $M$
of the functions $\Ga$ to be the pull-back of the Poisson brackets
of the $\ga$'s in the group $H^*$:
\begin{eqnarray}
\{  \Ga \x  \Ga  \}_M\ =\
\CP^*\ \{\ \ga \x  \ga\}_{H^*}\ =\
\eta^*(\Ga)\cdot \Ga\oti\ \Ga \label{EIxvii}
\end{eqnarray}
\proclaim Proposition.
Assuming this relation implies that the action defined
in eq.(\ref{EIxii}) is a representation of $\CH$ on the space of
functions $f$ on $M$:
\debut
(XY-YX)\cdot f =  <\Ga^{-1}\{f,\Ga,\}_M,\[X,Y\]_\CH>
\label{Erep}
\fin

This is proved by noticing that
\begin{eqnarray}
(XY-YX).f = <(Ad\Ga^{-1})_{12} \{f ,\eta_{12}(\Ga)\}_M ,X_1 Y_2 >
\nonumber
\end{eqnarray}
and using that $(\nabla^d_L\eta^*)(\ga)=
Ad\ga\cdot(\nabla^d_L\eta^*)(e)$ as follows from the cocycle
condition.
The converse is also true: requiring the relation (\ref{Erep})
implies the Poisson brackets (\ref{EIxvii}) for $\Ga$.

\bigskip
\noindent {\bf 1d- The groups $G$ and $G^*$.}
As a preparation to our study of dressing transformations, we apply
the previous results to a more specific situation.

{\it The group G}:
Let $\CG$ be a Lie algebra with a bilinear invariant form denoted by $tr$.
Let $G$ be its connected Lie group.
We choose an orthonormal basis in $\CG$; $t^a\in\CG$,
$tr(t^at^b)=\de^{ab}$ and $\[t^a,t^b\]=f^{ab}_c t^c$ where
$f^{ab}_c$ are the structure constants of $\CG$. We denote
by $\CC=\sum_a t^a\oti t^a$ the tensor Casimir in $\CG\oti\CG$.

Following Sklyanin \cite{Sk79}, the group $G$ is  equipped with a
Poisson structure as follows. Let $r^\pm\in\CG\oti\CG$,
$r^\pm=\sum_{a,b} r^\pm_{ab} t^a\oti t^b$, with $r^+_{12}=-r^-_{21}$
and $r^+-r^-=\CC$, be two solutions of the classical Yang-Baxter
equation:
\begin{eqnarray}
\[ r_{12}^\pm , r_{13}^\pm \]\ +\
\[ r_{12}^\pm , r_{23}^\pm \]\ +\
\[ r_{13}^\pm , r_{23}^\pm \]\ =\ 0 \label{ECi}
\end{eqnarray}
Eq.(\ref{ECi}) is an equation in $\CG\oti\CG\oti\CG$, and the indices on
$r^\pm$ refer to the copies of $\CG$ on which $r^\pm$ is acting.
Using the bilinear form $tr$ to identify the vector spaces $\CG^*$ and
$\CG$, the elements $r^\pm$ of $\CG\oti\CG$ can be mapped into elements
$R^\pm\in\CG\oti\CG^* \cong {\rm End}\CG$ defined by \cite{Se83}:
\begin{eqnarray}
R^\pm( X )\ =\ tr_2\({ r^\pm_{12} (1\oti X) }\)
\qquad;\qquad \forall\ X\in\CG \label{ECii}
\end{eqnarray}
Note that we have $R^+-R^-=Id$.
The Sklyanin  bracket is then defined for any functions
$f_1$ and $f_2$ on $G$ by
\begin{eqnarray}
\{ f_1,f_2\}_G\ =\ \sum_{a,b}\BL\
r^\pm_{ab} (\nabla^a_Rf_1)(\nabla^b_Rf_2)\ -\
r^\pm_{ab} (\nabla^a_Lf_1)(\nabla^b_Lf_2)\ \BR \nonumber
\end{eqnarray}
where $\nabla^a_{L,R}$ denote the left (right) invariant
vector fields corresponding to the elements $t^a\in\CG$. Taking as
functions the matrix elements in some representation of $G$, we have:
\begin{eqnarray}
\{\ x  \x  x\ \}_G\ =\ \[\ r^\pm\ ,\ x\oti x\ \]
\qquad;\qquad  x\in G \label{ECv}
\end{eqnarray}
In eq.(\ref{ECv}) we can choose to use either $r^+$ or $r^-$ as
the difference is the tensor Casimir. Comparing with the general
situation described above, we see that the cocycle $\eta$ is here a
coboundary $\eta(x)=r^\pm -Ad x.\ r^\pm $.

{\it The group} $G^*$: The Poisson bracket (\ref{ECv})
on $G$ induces a Lie algebra structure on $\CG^*$ by the formula
(\ref{EIvii}). Again, identifying the vector spaces ${\cal G}$ and ${\cal
G}^*$ by means of $tr$,  the brackets on $\CG^*$ read \cite{Se83}:
\begin{eqnarray}
\[\ X,Y\ \]_R\ &=&\ \[R^\pm(X),Y\]\ +\ \[X,R^\mp(Y)\]  \non
 	& =&\ \half\BL \[R(X),Y\]\ +\ \[X,R(Y)\] \BR \label{ECvi}
\end{eqnarray}
with $R=R^++R^-$. The Jacobi identity
for the commutators (\ref{ECvi}) follows from the classical Yang-Baxter
equation for $r^\pm$ . It also implies that $R^\pm$ are Lie algebra
homorphisms from $\CG^*$ to $\CG$. In particular,
$\CG_\pm={\rm Im}R^\pm$ are
subalgebras of $\CG$, and $r^\pm\in\CG_\pm\oti\CG_\mp$.
The group $G^*$ and the algebra $\CG^*$ are related
to a factorization problem in $G$ specified by the matrices $R^\pm$.
First consider $\CG^*$; because $R^+-R^-=Id$, any $X\in \CG$ admits a
unique decomposition as :
\begin{eqnarray}
X\ =\ X_+-X_-\qquad {\rm with}\qquad
X_\pm=R^\pm(X)=tr_2\({r^\pm(1\oti X)}\)
\label{ECvii}
\end{eqnarray}
In terms of the components $X_+$ and $X_-$, the commutator in
$\CG^*$ becomes:
\begin{eqnarray}
\[\ X,Y\ \]_R\ =\  \[X_+,Y_+\]\ -\ \[X_-,Y_-\] \nonumber
\end{eqnarray}
In particular, the plus and minus components commute in $\CG^*$.
We denote by $G^*$ the corresponding group.
By exponentiation, the group $G^*$ is made of the couples
$(g_-,g_+)$ with the composition law,
\begin{eqnarray}
(g_-,g_+)\cdot(h_-,h_+)=(g_-h_-,g_+h_+) ,\label{EClaw}
\end{eqnarray}
Just as $\CG\simeq \CG^*$ as vector spaces, we have $G\simeq G^*$
as manifolds: $(g_-,g_+)\in G^*\to g=g_-^{-1}g_+\in G$. Equivalently,
any elements $g\in G$ (in a neighbourhood of the
identity) admits a unique factorization as:
\begin{eqnarray}
g\ =\ g_-^{-1}\,g_+ \qquad {\rm with}\qquad (g_-,g_+)\in G^*
\label{ECix}
\end{eqnarray}
Notice the analogy between the definition of $G^*$ and the standard
notion of normal order in quantum field theory. The plus and minus
component of $g\in G$  in $G^*$ can be thought as the positive and
negative ``frequency part" of $g$. The product in $G^*$ is then the
product of the Wick ordered objects $g=:g_-^{-1}g_+:$ and
$h=:h_-^{-1}h_+$; i.e. $:g_-^{-1}g_+\ h_-^{-1}h_+: =
:(g_-h_-)^{-1} g_+h_+:$.

The group $G^*$ itself becomes a Poisson-Lie group if we
introduce on it the Semenov-Tian-Shansky Poisson bracket \cite{Se85}:
\begin{eqnarray}
\{g_+\x g_+\}_{G^*}\ &=&- \[r^\pm, g_+\oti g_+\] \nonumber \\
\{g_-\x g_-\}_{G^*}\ &=&- \[r^\mp, g_-\oti g_-\] \nonumber \\
\{g_-\x g_+\}_{G^*}\ &=&- \[r^-, g_-\oti g_+\] \nonumber \\
\{g_+\x g_-\}_{G^*}\ &=&- \[r^+, g_+\oti g_-\] \label{ECx}
\end{eqnarray}
or, for the factorized element $g=g_-^{-1}g_+$:
\begin{eqnarray}
\{g \x g\}_{G^*}\ &=& -(g\oti1)r^+(1\oti g)\ -\ (1\oti
g)r^-(g\oti1)\nonumber \\
&&+(g\oti g)r^\pm + r^\mp(g\oti g) .\label{ECxi}
\end{eqnarray}
The multiplication in $G^*$ is a Poisson map for the brackets
(\ref{ECx}). The group $G^*$ is therefore a Poisson-Lie group. Notice
that, as it should be, the Lie algebra structure induced
by the Poisson brackets (\ref{ECxi})
on the dual of $\CG^*$, i.e. on $\CG$,
is the original structure on $\CG$.

{\it The action of $G^*$ on $G$.} This is a part of the multiple
relations between $G$ and $G^*$ forming the theory of the
classical double \cite{Dr86,Se85}.
We consider the  action of
$G^*$ on $G$ defined, for any $(g_-,g_+)\in G^*$, as follows:
\begin{eqnarray}
(g_-,g_+)\in G^*,~ x\in G\ \to x^g\ =\ (x g x^{-1})_\pm\, x\,
g^{-1}_\pm\ \in G
\qquad {\rm with}\qquad g=g_-^{-1}g_+ \label{ECxii}
\end{eqnarray}
The two signs give the same result for $x^g$.
First, we have to justify that this is really an action of $G^*$ on $G$,
i.e. we have to prove that the composition law of the transformations
eq.(\ref{ECxii}) is the same as the composition law in $G^*$. Therefore,
consider two elements $g=g_-^{-1}g_+$ and $h=h_-^{-1}h_+$ and
transform successively  an element $x\in G$: $x\to x^g\to (x^g)^h$;
we have \cite{Se85,AvBe}:
\begin{eqnarray}
x^g\ =\ \Th^g_\pm\ x\ g^{-1}_\pm
\qquad &{\rm with}&\qquad \Th^g_\pm = \({x g x^{-1}}\)_\pm \label{ECxiv} \\
(x^g)^h\ =\ \Th^{hg}_\pm\ x^g\ h^{-1}_\pm
\qquad &{\rm with}&\qquad \Th^{hg}_\pm = \({x^g h x^g\,^{-1}}\)_\pm
\nonumber
\end{eqnarray}
The factorization of $(x^ghx^g\,^{-1})$ can be written as follows:
\begin{eqnarray}
(\Th^{hg}_-)^{-1}\Th^{hg}_+\ \equiv x^ghx^g\,^{-1}\ =\
\Th^g_-\,x\, (h_-g_-)^{-1}(h_+g_+)\, x^{-1}\,{ \Th^g_+}^{-1} \nonumber
\end{eqnarray}
or, equivalently,
\begin{eqnarray}
 \Theta_\pm^{hg} \Theta_\pm^g =\({ x\, (h_-g_-)^{-1}(h_+g_+)\, x^{-1}
}\)_\pm\
 \nonumber
\end{eqnarray}
Inserting this formula into eq.(\ref{ECxiv}) proves that the multiplication
law for the dressing transformations is the same as in $G^*$.

One of the main property of this action is that it is a Lie-Poisson
action of $G^*$ on $G$ if the groups $G$ and $G^*$ are equipped with
the Poisson structures defined in eqs.(\ref{ECv}) and (\ref{ECx});
i.e. we have:
\debut
\{x^g \x  x^g\}_{G\times G^*}\ =\
\[r^\pm\ ,\ x^g\oti x^g\] \nonumber
\fin
This was proved in
\cite{Se85} using the classical double.
We will give another proof of this fact below.

The infinitesimal form of eq.(\ref{ECxii}) is for any $X\in \CG$,
with $X=X_+-X_-$:
\begin{eqnarray}
\de_X\ x\ =\ Y_\pm\ x - x\ X_\pm
\qquad {\rm with}\qquad Y_\pm = (x X x^{-1})_\pm \label{ECxiii}
\end{eqnarray}
\proclaim  Proposition.
The non-Abelian Hamiltonian of
the transformations (\ref{ECxii}),
which is an element of $G^{**}\cong G$, is the
group element itself, i.e.:
\begin{eqnarray}
\de_X\, x\ =\ tr_2\({ (1\oti Xx^{-1})\ \{x\x x\}_G }\)
\quad \forall X\in\CG,\ \forall x\in G \label{ECxvii}
\end{eqnarray}
\par
Indeed, from eq.(\ref{ECv}) we have,
$\forall\ X\in \CG$:
\begin{eqnarray}
tr_2\({ (1\oti Xx^{-1}) \{x\x x\}_G }\)
&=& tr_2\({ r^\pm(1\oti xXx^{-1}) }\)\, x -
x\, tr_2\({ r^\pm(1\oti X)}\)\cr
&=& (xXx^{-1})_\pm\, x - x\, X_\pm\ =\ \de_X\, x \label{ECxviii}
\end{eqnarray}
In the last step we used eq.(\ref{ECii}).
It is a remarkable fact that there exists
a non-Abelian Hamiltonian since the group $G$ with
the Sklyanin bracket is not a symplectic manifold.\\
The existence of a non-Abelian Hamiltonian for these
transformations gives a two-line proof of their
Lie Poisson property. Indeed, using the Jacobi identity
for $\{,\}_G$, we have:
\debut
\de_X\{x_1,x_2\}_G\ &-&\ \{\de_Xx_1,x_2\}_G\ -\ \{x_1,\de_X x_2\}_G\non
&=&\ {\rm tr}_0\BL X_0 \BBL x_0^{-1}\{x_1,x_0\}_G\ ,\
x_0^{-1}\{x_2,x_0\}_G\ \BBR\BR \nonumber
\fin
where we put $x_0=x\oti1\oti1$, $x_1=1\oti x\oti1$
and $x_2=1\oti1\oti x$; it is just eq.(\ref{EIx}).

\bigskip
{\bf 1e- The dressing transformations.}
Dressing transformations are special symmetries of soliton
equations which admit a Lax representation \cite{ZaSh79,DaJiKaMi81}.
Namely, consider a set of non-linear differential equations for some
fields $\phi(x,t)$, which can be written
as a zero-curvature condition:
\debut
\BBL \CD_\mu\[\phi\] , \CD_\nu[\phi] \BBR = 0\label{Edri}
\fin
for a connexion $\CD_\mu\[\phi\]=\d_\mu - A_\mu\[\phi\]$ depending
on the fields $\phi$. The connexion $A_\mu\[\phi\]$ is usually
called the Lax connexion; it belongs to some Lie algebra $\CG$
and its form depends on the soliton equations. The zero-curvature
eq.(\ref{Edri}) is the compatibility condition for an auxiliary
linear problem:
\debut
(\d_\mu\ -\ A_\mu ) \Psi(x,t)\ =\ 0 \label{Edrii}
\fin
where the wave function $\Psi(x,t)$ takes values in the group $G$.
Due to the zero-curvature condition, the Lax connexion is a pure
gauge:
$A_{\mu}\ =\ \BL\d_{\mu}\Psi\BR\ \Psi^{-1}$. The wave function $\Psi(x,t)$ is
defined up to a right multiplication by
a space-time independent group element. This freedom is fixed
by imposing a normalization condition on $\Psi$; e.g. $\Psi(0)=1$.

Suppose now that the soliton equations (\ref{Edri}) admit
an Hamiltonian formulation \cite{Fa82,FaTa86} and moreover that the Poisson
brackets of the components of the Lax connexion (which are
deduced from those of the fields $\phi$) give rise to the
Sklyanin brackets for the function $\Psi(x,t)$:
\begin{eqnarray}
\Bigl\{\ \Psi(x)\ \x  \Psi(x)\ \Bigr\}\ =\
\BBL\ r^{\pm}\ ,\ \Psi(x)\oti\Psi(x)\ \BBR \label{extra}
\end{eqnarray}
where $r^\pm\in\CG\oti\CG$ are solutions of the classical
Yang-Baxter equation. As  we explained in the previous section,
the matrices $r^\pm$ can be used to define two subalgebras
$\CG_\pm$ of $\CG$ and the associated factorization problems,
either in the algebra $\CG$: $X=X_+-X_-$ or in the group
$G$: $g=g_-^{-1}g_+$.

For any $g=g_-^{-1}g_+\in G$,
a dressing transformation  consists in transforming the
variables $\Psi(x,t)$ to the variables $\Psi^g(x,t)$ defined by
\begin{eqnarray}
\Psi^g\  =\ \Th_\pm\ \Psi\ g_\pm^{-1} \label{EIIi}
\end{eqnarray}
where the group elements $\Th_\pm$ are solutions of the factorization problem:
\begin{eqnarray}
\Th^{-1}_-\ \Th_+\ =\ \Psi\ g\ \Psi^{-1} \label{EIIii}
\end{eqnarray}
Notice that the two signs give the same result in eq.(\ref{EIIi}).
Also this transformation preserves the normalization condition $\Psi(0)=1$.
The action on the wave function eq.(\ref{EIIi}) induces a gauge
transformation on the
Lax connexion : $A_\mu = (\d_\mu\Psi)\ \Psi^{-1}$ is transformed into
$A^g_\mu = (\d_\mu\Psi^g)\ \Psi^g\,^{-1}$ with:
\begin{eqnarray}
A^g_\mu\ =\ (\d_\mu\Th_\pm)\Th_\pm^{-1}\ +\
\Th_\pm A_\mu \Th^{-1}_\pm \label{EIIiii}
\end{eqnarray}
In general \cite{ZaSh79,Se85}, and as we will illustrate by examples in the
following sections, the factorization problem is constructed
in such a way that the form of the Lax connexion is preserved
by the dressing transformations. This is the main property of
these transformations. It implies that they form a (usually
called ``hidden") symmetry group of the soliton equations.

The transformations (\ref{EIIi}) are similar to the action of
$G^*$ on $G$ we considered in previous sections. Therefore, \\
---(i) the commutation relations of the dressing transformations
are those of $G^*$ (in particular the plus and minus components
commute),\\
---(ii) the action of $G^*$ on the phase space
induced by eq. (\ref{EIIi}) is a Lie Poisson action if the
dressing group is equipped with the Semenov-Tian-Shansky's bracket.\\
---(iii) the non-Abelian Hamiltonian of the
Lie-Poisson action (\ref{EIIi}) is the monodromy matrix.
This matrix is $T(L)=\Psi(L)\Psi^{-1}(0)$, with
$0$ and $L$ the two boundary values of the coordinate $x$.
Using the ultralocality property, its Poisson
bracket with the wave function is:
\debut
\Bigl\{\ \Psi(x) \x T(L)\ \Bigr\}\ =\
\BL 1\oti T(L)\Psi^{-1}(x)\BR
\BBL\ r^{\pm}\ ,\ \Psi(x)\oti\Psi(x)\ \BBR \nonumber
\fin
Therefore, in the same way as in eq.(\ref{ECxviii}), we have:
\debut
tr_2\BL1\otimes XT^{-1}(L)
\Bigl\{ \Psi(x) \x T(L) \Bigr\}\BR\ &=&\
\BL\Psi(x)X\Psi^{-1}(x)\BR_\pm \Psi(x) -
\Psi(x) X_\pm \non
&=& \de_X\,\Psi(x) \nonumber
\fin
for any $X\in\CG$. This proves that $T(L)$ is the non-Abelian
Hamiltonian. We will illustrate this relation in various ways in
the following.

\bigskip
\section{Example 1: Dressings in the Toda field Theories.}

In this section, we apply the previous results
to describe the dressing
transformations in Toda field theories.
The dressing transformations in the Toda field
theories reflect an invariance of these models
under a semi-classical version of the quantum algebras $\CU_q(\CG^*)$.
It provides a simple explanation of the occurence of quantum groups in
conformal
field theory. Some of these results were announced in \cite{BaBe91}.

\bigskip
\noindent {\bf 2a- The Toda field theories.}
Let us first fix the notations. Let $\CG$ be a semi-simple Lie
algebra equipped with its Killing form denoted by $(,)=tr$. We denote by $G$
its Lie group. Let $\CH$ be a Cartan subalgebra of $\CG$,
$\CG=\CN_-\oplus\CH\oplus\CN_+$ the Cartan decomposition,
$G=N_-\ H\ N_+$ the corresponding Gaussian decomposition and
$B_\pm=HN_\pm$ the Borel subgroup. We denote
by $(e^-_i,\ \alv_i, \ e^+_i)$, $i=1,\cdots,{\rm rank}\CG$,
the generators associated to the simple roots $\al_i$
of $\CG$, $\al_i\in\CH$. They satisfy:
\begin{eqnarray}
\BBL\ \alv_i\ ,\ \alv_j\ \BBR\ &=& 0 \nonumber \\
\BBL\ \alv_i\ ,\ e^{\pm}_j\ \BBR\ &=& \pm\ \al_j(\alv_i)\ e^{\pm}_j
\nonumber \\
\BBL\ e^+_i\ ,\ e^-_j\ \BBR\ &=& \de_{ij}\ \alv_i \nonumber
\end{eqnarray}
and $\BL ad\ e^{\pm}_i \BR^{1-a_{ij}}\ e^{\pm}_j = 0$
for $i\not= j$, with $a_{ij} = \al_j(\alv_i)$ the Cartan matrix of
$\CG$. The algebra $\CG$ can be equipped with a natural gradation,
sometimes called the principal gradation. It is specified by
setting $deg(\alv_i)=0$; $deg(e^{\pm}_i)=\pm1$.

We denote by $(x,t)$ the space-time coordinates and by
$x^{\pm}= x\pm t$, the light-cone coordinates.  We consider
the model on a cylinder of radius $L$ (therefore, $x$ goes from
$0$ to $L$). Let $\Phi(x,t)$
be a field taking values in the Cartan subalgebra $\CH$.
The Toda field equations for $\Phi(x,t)$ are:
\begin{eqnarray}
\d^{\mu}\d_{\mu}\ \Phi(x,t)\ =\
\half\ \sum_i\ \alv_i \exp\BBL\ 2\al_i\BL\Phi(x,t)\BR\BBR \label{ei}
\end{eqnarray}
The sum in eq.(\ref{ei}) is over the simple roots of $\CG$.
As is well known, eq.(\ref{ei}) admits a Lax representation. The
Toda equations of motion (\ref{ei}) are equivalent to the zero curvature
condition,
$\BBL \CD_{\mu} , \CD_{\nu} \BBR = 0$
for the Lax connexion, $\CD_{\mu}=\d_{\mu}-A_{\mu}$, defined by,
\begin{eqnarray}
\CD_{x^+}\ &=& \d_{x^+}\ +\ \d_{x^+}\Phi\ +\ e^{ad\Phi}\ \CE_+
\nonumber \\
\CD_{x^-}\ &=& \d_{x^-}\ -\ \d_{x^-}\Phi\ +\ e^{-ad\Phi}\ \CE_-
\label{eiii}
\end{eqnarray}
with $\CE_{\pm}=\sum_i e^{\pm}_i$. Note that $deg(\CE_{\pm})=\pm 1$.
Let, as in the previous section, $\Psi(x,t)$ be the normalized wave function;
it is specified by $(\d_\mu-A_\mu)\Psi(x,t)=0$ and the normalization
condition $\Psi(0)=1$.

The Toda field theories are Hamiltonian systems. The Poisson
brackets between the field $\Phi(x,t)$ and its conjugate momentum
$\Pi(x,t)$ are introduced in the usual way. As is well known
it implies the following Poisson brackets for
the $G$-valued function $\Psi(x,t)$:
\begin{eqnarray}
\Bigl\{\ \Psi(x)\ \x \ \Psi(x)\ \Bigr\}\ =\
\BBL\ r^{\pm}\ ,\ \Psi(x)\oti\Psi(x)\ \BBR \nonumber
\end{eqnarray}
The matrices $r^{\pm}$ are given by \cite{BeDr84,OlTu83}:
\begin{eqnarray}
r^{\pm}\ =\ \pm \BL \sum_i\ H_i\oti H_i +
2\sum_{\al>0}\ E_{\pm \al}\oti E_{\mp \al}\ \BR \label{eviii}
\end{eqnarray}
with $H_i$ an orthonormalized basis of $\CH$ and
$E_{\pm \al}$ the normalized Chevalley generators.
The sum $\sum_{\al>0}$ is over all the positive roots
of the algebra $\CG$. The matrices
$r^{\pm}$ are solutions of the classical Yang-Baxter equation (\ref{ECi}).

\bigskip
\noindent {\bf 2b- Dressing transformations in the Toda models.}
The dressing transformations are associated to a factorization
problem in the group $G$. For the Toda field
theories, this factorization is
specified by the classical $r^\pm$ - matrices given in  eq.(\ref{eviii}).
It can be described as follows.
Any element $g\in G$ admits a unique decomposition
as $g=g_-^{-1}\ g_+$ with $g_{\pm}\in B_{\pm}=HN_\pm$ and such that
$g_-$ and $g_+$ have inverse components on the Cartan torus.
In practice it is given by half splitting the Gaussian decompostion of $g$.
The infinitesimal version of this factorization problem consists
in decomposing any element $X\in\CG$ as $X=X_+-X_-$ with
$X_\pm\in(\CH\oplus\CN_\pm)$ such that $X_+$ and $X_-$ have
opposite components on $\CH$.

The dressing transformations are defined as in eqs.(\ref{EIIi})
and (\ref{EIIii}): $\Psi(x)\to \Psi^g(x)=\Th_\pm \Psi(x) g_\pm^{-1}$
with $\Th_-^{-1}\Th_+=\Psi(x)g\Psi^{-1}(x)$.
They induce gauge transformations on the
Lax connexion : $A_\mu\ =\ (\d_\mu\Psi)\ \Psi^{-1}$ is transformed into
$A^g_\mu\ =\ (\d_\mu\Psi^g)\ \Psi^g\,^{-1}$.
The factorization problem described above is cooked up such that the
form of the Lax connexion
is preserved by these transformations.
The proof of this statement essentially relies on the fact that
the gauge transformations (\ref{EIIiii}) can be implemented using either
$\Th_-$ or $\Th_+$.
One first shows that the degrees of the components of the Toda
Lax connexion are preserved by the dressing and then, one
verifies that the connexion can be written as in eq.(\ref{eiii}).

The (non-local) gauge transformations eq.(\ref{EIIiii}) of the Lax
connexion induce
transformations of the Toda fields $\Phi(x,t)$.
Because the form of the Lax connexion is preserved by these
transformations, the induced  actions
map a solution of the Toda equations $\Phi(x,t)$ into
another solution $\Phi^g(x,t)$ (which, in general,
possess non-trivial topological numbers).
The transformations of the Toda fields can be described as follows.
Factorize $\Th_\pm$ as :
\begin{eqnarray}
\Th_\pm\ =\ (\Psi g \Psi^{-1})_\pm\ =\ K^g_\pm\ M^g_\pm \nonumber
\end{eqnarray}
with $M^g_\pm\in N_\pm$ and $K^g_\pm\in H$. According to the
factorization problem, the components of $\Th_-$ and $\Th_+$
on the Cartan torus are inverse: $K^g_-K^g_+=1$.
Put $K^g_\pm\ =\ \exp \Delta^g_\pm$ with $\Delta^g_\pm\in\CH$:
$\De^g_-+\De^g_+=0$. Then,
by looking at the exact expression of the transformed
Lax connexion $A_{\mu}^g$, one deduces that:
\begin{eqnarray}
\Phi^g\ =\ \Phi - \De^g_+\ =\ \Phi + \De^g_- \label{exiii}
\end{eqnarray}
By construction, if $\Phi$ is a solution of the Toda equations,
so is the dressed field $\Phi^g$. The relation between $\Phi^g$ and
$\Phi$ is non-local because $\De^g_{\pm}$
are expressed in a non-local way in terms of $\Phi$.
We have: $\exp(\pm 2\lam(\De^g_\pm))=\bra{\lam}\Psi g\Psi^{-1}\ket{\lam}$,
for any highest weight $\lam$.

We explicitly checked in \cite{BaBe91} that the action
defined in eq.(\ref{EIIi}) is a Lie Poisson action provided that the group
$G^*$ is equipped with the Semenov-Tian-Shansky Poisson
brackets (\ref{ECx}).

\bigskip
\noindent {\bf 2c- Dressings of the chiral fields.}
If $\CG$ possesses highest weight representations,
the Toda field equations admit a remarkable chiral splitting.
Requiring that the Kac-Moody algebra possesses integrable highest
weight representations selects the finite dimensional Lie algebras
or the affine algebras but excludes the loop algebras.
Let us denote by $\rho$ a highest
weight vector representation of $\CG$ acting on the linear space $V$,
and by $\ket{\lam}$ its highest weight vector. To any representation
$\rho$ we associate two fields, $\xib(x,t)$ and $\xi(x,t)$,
taking values in $V$ and in its  dual $V^*$. They are defined by:
\begin{eqnarray}
\xi(x)\ &=& \bra{\lam}\ \BL\ e^{-\Phi(x)}\ \Psi(x)\ \BR  \label{ev a} \\
\xib(x)\ &=& \BL\ \Psi^{-1}(x)\ e^{-\Phi(x)}\ \BR\ \ket{\lam}
\label{ev b}
\end{eqnarray}
$\xi(x)$ is a line vector, $\xib(x)$ is a column vector.
The Toda field equations imply that these fields are chiral:
$\d_{x^-}\xi=\d_{x^+}\xib=0$. The Toda fields $\Phi(x)$
can be reconstructed from the fields $\xi(x)$ and $\xib(x)$:
\begin{eqnarray}
\exp\BL\ -2\lam(\Phi)(x)\ \BR\ =\ \xi(x)\ .\ \xib(x) \nonumber
\end{eqnarray}
The Poisson brackets between the fields $\xi(x)$ and $\xib(x)$ are
derived from the Sklyanin brackets (\ref{extra}) as in
\cite{Ba88b}.  They are the following semi-classical exchange relations:
\begin{eqnarray}
\Bigl\{\ \xi(x)\x \xi(y)\ \Bigr\}\ &=& -\xi(x) \otimes \xi(y)r^\pm
\qquad\qquad\qquad;\ {\rm for}\ x\ {^> _<}\ y \nonumber\\
\Bigl\{\ \xib(x)\x \xib(y)\ \Bigr\}\ &=&- r^\mp \xib(x) \otimes \xib(y)
\qquad\qquad\qquad;\ {\rm for}\ x\ {^> _<}\ y \nonumber\\
\Bigl\{\ \xi(x)\x \xib(y)\ \Bigr\}\ &=&
\BL\xi(x)\oti 1\BR\ r^-\ \BL 1 \oti\xib(y)\BR
\qquad \qquad;\ \forall\ x,y \nonumber \\
\Bigl\{\ \xib(x)\x \xi(y)\ \Bigr\}\ &=&
\BL 1\oti\xi(y)\BR\ r^+\ \BL\xib(x)\oti 1\BR
\qquad \qquad;\ \forall\ x,y \nonumber
\end{eqnarray}

\proclaim  Proposition.
The action on the chiral fields
$\xi(x,t)$ and $\xib(x,t)$ induced by the dressing
transformation is remarkably simple:
\begin{eqnarray}
\xi^g(x,t)\ &=& \xi(x,t)\ .\ g_-^{-1}\nonumber \\
\xib^g(x,t)& =& g_+\ .\ \xib(x,t)\label{exv}
\end{eqnarray}
It gives another formula for the dressed Toda fields:
\debut
\exp\BL -2\lam(\Phi^g)(x) \BR\ =\ \xi(x).g. \xib(x) \nonumber
\fin
\par
The transformation properties of the
chiral fields $\xi(x,t)$ and ${\bar \xi}(x,t)$
shows clearly that the plus and minus components in $G^*$ commute.
Note that the $g_{\pm}$ components of $g$ act separately on the two
chiral sectors.
The proof of eq.(\ref{exv}) is very simple:
\begin{eqnarray}
\xi^g\ &=& \bra{\lam}\ \BL e^{-\Phi^g}\ \Psi^g \BR \nonumber \\
&=& e^{-\lam(\Phi^g)}\ \bra{\lam}\ \BL \Th_-\ \Psi\BR\ g_-^{-1}\non
&=&\ e^{-\lam(\Phi^g-\Phi-\De^g_-)}\
\bra{\lam}\ \BL e^{-\Phi}\ \Psi\BR\ g_-^{-1}\nonumber \\
&=& \xi\ .\ g_-^{-1} \label{exvi}
\end{eqnarray}
In eq.(\ref{exvi}) we implemented the transformation using
$\Th_-$ and $g_-$, and we used the property that $\ket{\lam}$
is a highest weight vector and the relation (\ref{exiii}).
The transformation law eq.(\ref{exv}) of the chiral fields probably
provides the simplest way to derive the Poisson structure
(\ref{ECx}) of the dressing group.

\bigskip
\noindent {\bf 2d- Non-local charges and the monodromy matrix.}
Let us first check that the monodromy
matrix $T=\Psi(L)$ is the non-Abelian Hamiltonian generating the
dressing transformations. The Poisson brackets between the chiral
fields and the monodromy matrix are:
\begin{eqnarray}
\{\xi(x) \x T\}\ &=& -\xi(x)\oti T\ r^-\nonumber \\
\{\xib(x) \x T\}\ &=&  (1\oti T) r^+ (\xib(x)\oti 1)\nonumber
\end{eqnarray}
Therefore, by a direct computation, we find, for any $X\in\CG$:
\begin{eqnarray}
tr_2\({ 1\oti XT^{-1}\ \{\xi(x)\x T\} }\)\
&=& -\xi(x) X_-\ =\ \de_X\xi(x)
\nonumber \\
tr_2\({ 1\oti XT^{-1}\ \{\xib(x)\x T\} }\)\
&=&  X_+\xib(x)\ =\ \de_X\xib(x)\label{EIIvi}
\end{eqnarray}
Eqs.(\ref{EIIvi}) are the infinitesimal actions of the dressings (\ref{exv})
by $g\simeq 1 + X_+ - X_-$.\\
As explained in Sect. 1,
the Poisson algebra of the charges which generate the dressing
follow from the Poisson brackets of the monodromy matrix:
\begin{eqnarray}
 \{ T \x T \} = [r^\epsilon , T\otimes T ] \nonumber
 \end{eqnarray}
To give a more precise description of this algebra,
we  introduce a more adequate parametrisation of the monodromy matrix $T$. This
parametrization is obtained by factorizing $T=T_-^{-1}T_+$ with:
\begin{eqnarray}
T_\pm = D^{\pm 1} M_\pm~~~~~~ \mbox{with}~~~~~ D \in \exp{\cal H},
 ~~ M_\pm \in \exp{\cal N}_\pm
\nonumber
\end{eqnarray}
The Poisson brackets for the matrix $T$  then become
\begin{eqnarray}
\{ D \x D \} &=& 0 \nonumber \\
\{ M_\pm \x D \} (M_\pm^{-1} \otimes D^{-1} ) &=& \pm {1\over 2}
\sum_i ~ \left( M_\pm H_i M_\pm^{-1} \otimes H_i - H_i \otimes H_i \right)
\nonumber \\
 \{ M_+ \x M_- \} &=& 0 \nonumber \\
 \{ M_\pm \x M_\pm \}(M_\pm^{-1} \otimes M_\pm^{-1}) &=& \left(
 r^\epsilon - M_\pm \otimes M_\pm r^\epsilon M_\pm^{-1}
\otimes M_\pm^{-1} \right)
 \Big\vert_{\CN_\pm \oti \CN_\pm } \nonumber
 \end{eqnarray}
Notice that the Poisson brackets between $M_+$ and $M_-$ are zero and
that the matrix $D$ is a generating function for charges in involution.
The matrices $T_\pm$ are related to the charges $Q_i$ and $Q_{\pm\al}$ by:
\debut
D^2 &=& \exp\left( \sum_i Q_i H_i\right) \nonumber\\
M_\pm &=& \exp \left( 2 \sum_{\al>0} Q_{\pm \al} E_{\pm\al}\right)
\nonumber
\fin
\proclaim  Proposition.
The dressing Poisson algebra is generated by the charges
$Q_i^\vee=(\al_i^\vee,Q)=\sum_j (\al_i^\vee,H_j)Q_j$ and $Q_{\pm\al_i}$,
where $\al_i$ are the simple roots.
The only relations among these generators are:
\debut
\{Q_i^\vee, Q_j^\vee\} &=& 0 \non
\{Q_i^\vee, Q_{\pm\al_j} \} &=& \pm a_{ij} Q_{\pm\al_j} \nonumber \\
\{ Q_{\pm\al_i}, Q_{\mp\al_j}\}&=&0 \label{EPal}
\fin
plus the deformed Serre-like relations:
\debut
\sum_{0\leq 2k \leq 1-a_{ij}}\ B_{ij}^{2k} Q_{\pm\al_i}^{2k}
\underbrace{\{Q_{\pm\al_i},\{\cdots,\{Q_{\pm\al_i}}_{1-a_{ij}-2k~~ factors}
,Q_{\pm\al_j}\} \cdots \} =0 \label{EPalg}
\fin
where the $B_{ij}$-coefficients are given by the generating
function $B_{ij}(x)$:
\debut
B_{ij}(x)=\sum_{0\leq 2k\leq 1-a_{ij}} B_{ij}^{2k} x^{2k}
=\prod_{0\leq 2p\leq -a_{ij}}\({1-\(\al_i,\al_j+p\al_i\)^2\,x^2}\)
\nonumber
\fin
\par
 For the proofs see \cite{Ba88a}. Eqs.(\ref{EPalg}) are exactly
the semi-classical limits of the quantum Serre relations.
As it should be this Poisson algebra is the semi-classical limit of
$\CU_q(\CG^*)\simeq \CU_q(\CB_+)\times\CU_q(\CB_-)$, which is not
quite the same as the semi-classical limit of $\CU_q(\CG)$.

The action of these charges on the chiral fields is described as follows:
\debut
\{\xi(x),Q_i\}&=& \xi(x)\cdot H_i \nonumber\\
\{\xi(x),Q_{\al_i}\} +  \(\xi(x)\cdot H_{\al_i}\) Q_{\al_i}
&=&\ \xi(x)\cdot E_{-\al_i}\nonumber\\
\{\xi(x),Q_{-\al_i}\}&=& 0 \nonumber
\fin
Similarly with the other chirality. The extra piece in the bracket between
the charges $Q_{\al_i}$ and the chiral fields arises from the non-locality
of the charges: it reflects the semi-classical exchange relations between the
chiral fields and the non-local currents.

The fields $\xi(x,t)$ and ${\bar \xi}(x,t)$ are
the classical analogues of the field multiplets which
transform covariantly under the action of the quantum group
symmetry in the $W_\CG$-invariant conformal field theories,
(theories which include the minimal conformal models for $\CG=su(2)$)
\cite{FeWi91,GoSi91}.

\section{Example 2: The Heisenberg Model.}

As a second example, we consider the Heisenberg chain. We restrict ourselves
to the isotropic case, although the same considerations also apply to the non
isotropic cases. Besides being
one of the simplest and most studied model, our present interest lies
in the fact  that the group of dressing transformations is infinite
dimensional. We will show that the Poisson algebra of these transformations
form a Yangian. Recall that
according to Drinfeld \cite{Dr86}, the Yangian $Y({\cal G})$ is
an associative algebra with unity generated by the elements
$Q^0_i$ and $Q^1_i$ with the defining relations
\debut
\[Q^0_i ,Q^0_j\] &=& f_{ijk} Q^0_k \nonumber \\
\[Q^0_i,Q^1_j\] &=& f_{ijk} Q^1_k \nonumber \\
\[Q^1_i,\[Q^1_j,Q^0_k\]\]-\[Q^0_i,\[Q^1_j,Q^1_k\]\] &=&
A^{ijk}_{lmn}\{Q^0_l,Q^0_m,Q^0_n\}_{sym} \non
\[\[Q^1_i,Q^1_j\],\[Q^0_k,Q^1_l\] \]
&+&\[\[Q^1_k,Q^1_l\],\[Q^0_i,Q^1_j\]\] \non
=(A^{ijr}_{stu}f_{klr}
&-&A^{klr}_{stu}f_{ijr})\{Q^1_s,Q^0_t,Q^0_u\}_{sym}\label{EHi}
\fin
where $f_{ijk}$ are the structure constants of ${\cal G}$,
$A^{ijk}_{rst}=\frac{1}{24}f_{ira}f_{jsb}f_{ktc}f^{abc}$
and $\{x_1,x_2,x_3\}_{sym}=\sum_{i\not= j\not= k}x_ix_jx_k$.
It is a Hopf algebra with comultiplication
\debut
\Delta Q^0_i &=& Q^0_i\oti 1 + 1\oti Q^0_i \non
\Delta Q^1_i &=& Q^1_i\oti 1 + 1\oti Q^1_i -\half
f_{ijk} Q^0_j\oti Q^0_k \label{EHii}
\fin
The antipode and the counit can be deduced from this comultiplication.
The Yangian $Y(\CG)$ acts on the fields $\Phi$ through the adjoint action,
which for the generators $Q^0_i$ and $Q^1_i$ is given by:
\debut
(AdQ^0_i)\cdot\Phi &=& [Q^0_i,\Phi] \non
(AdQ^1_i)\cdot\Phi &=& [Q^1_i,\Phi] +
\half f_{ijk} [Q^0_j,\Phi] Q^0_k \label{EHadj}
\fin

Our purpose is to obtain the semiclassical version of the above
formulae by investigating the group of dressing transformations in the
Heisenberg model.

\bigskip
\noindent {\bf 3a- Definition of the model.}
We start with the definition of the model. Consider a spin
variable $S(x)$:
\debut
S(x)=\sum_{i=1}^3 S^i(x) \sigma_i\quad
{\rm with}\quad \sum_{i=1}^3 S^i(x)^2=s^2 \nonumber
\fin
where the $\sigma_i$ are the Pauli matrices with $[\sig_i,\sig_j]=
2i\ep_{ijk}\sig_k$ and $tr(\sig_j\sig_k)=2\de_{jk}$.
Introduce the Poisson bracket
\debut
\{ S^i(x),S^j(y)\} = \epsilon^{ijk} S^k(x)\delta(x-y) \nonumber
\fin
The Hamiltonian is
\debut
I_1=-{1\over 4}\int_0^L dx~ tr(\partial_x S \partial_x S) \nonumber
\fin
The equations of motion deduced from this Hamiltonian read
\debut
\d_t S
=-{\textstyle{i\over 2}} [ S ,\partial_x^2 S ]={\textstyle{i\over 2}}
\partial_x [S_x ,S]  \label{EHv}
\fin
Notice that these equations are the conservation laws
for a $su(2)$-valued current.
The integrability of the Heisenberg model relies on the fact
that the equations (\ref{EHv})
can be written as a zero curvature condition
\debut
\BL\partial_x + A_x \BR\Psi(x,\la)& =&0\non
\BL\partial_t + A_t\BR\Psi(x,\la) &=&0\nonumber
\fin
with Lax connexion:
\debut
A_x& =& {i\over \lambda}S(x) \non
A_t& =&- {{2is^2}\over \lambda^2} S(x) +
{1\over{2\lambda}}[S(x),\partial_x S(x)]\label{EHvii}
\fin
The Lax connexion (\ref{EHvii}) is an element of the $su(2)$ loop
algebra ${\tilde {su(2)}}=su(2)\otimes C\[\la,\la^{-1}\]$.
An important ingredient is the transport matrix $T(x,\lambda)$;
it is defined as
\debut
\Psi(x,\lambda)&=& T(x,\lambda)\Psi(0,\lambda) \non
T(x,\lambda)&=&P \exp\left[-\int_0^x A_x(y,\lambda) dy\right]
\label{EHviii}
\fin
The monodromy matrix $T(\lambda)$ is simply $T(L,\lambda)$.
As is well known, the importance of the monodromy matrix lies in
the fact that one can calculate the Poisson bracket of its  matrix
elements. One finds \cite{Sk79}:
\debut
\{ T(\lambda) \x T(\mu)\}={\textstyle{1\over 2}}[r(\lambda,\mu),
T(\lambda)\otimes T(\mu)]
\label{EHix}
\fin
where
\debut
r(\lambda ,\mu)={{\sum_i \sigma_i \otimes \sigma_i}\over{\lambda -\mu}}
\label{EHx}
\fin
 From this result, it follows that $tr(T(\lambda))$ generates
quantities in involution.

\bigskip
\noindent {\bf 3b- The transfer matrix and the local conserved charges.}
 From its definition, $T(x,\lambda)$ is analytic in $\lambda$
with an essential singularity at $\lambda =0$. From eq.(\ref{EHviii}),
we can easily find an expansion around $\lambda =\infty $:
\debut
T(x,\lambda )= 1 -{\textstyle{i\over \lambda}}\int_0^x dy S(y)
 -{\textstyle{1\over \lambda^2}}\int_0^x dy S(y) \int_0^y dz S(z) \cdots
\nonumber
\fin
This development in $1/\la$ has an infinite radius of convergence.
We will study it later in more details in relation with the
dressing transformations.

To find the structure of $T(x,\lambda)$ around $\lambda =0$ is more
delicate but important as it provides the local conserved charges in
involution.
The main point \cite{Fa82} is to notice that  there
exists a ${\it local}$ gauge transformation,
${\it regular}$ at $\lambda =0$, such that
\debut
T(x,\lambda)= g(x) D(x) g^{-1}(0) \label{EHxii}
\fin
where $D(x)$ is a diagonal matrix: $D(x)=\exp(id(x)\sig_3)$.
We can choose $g$ to be unitary, and,
since $g$ is defined up to a diagonal matrix, we can
require that it has a real diagonal:
\debut
g=\frac{1}{(1+v{\bar v})^\half}
\pmatrix{1 & v \cr -{\bar v} & 1\cr} \nonumber
\fin
The differential equation for $T$ becomes a differential
equation for $g$ and $d$:
\debut
g^{-1}\partial_x g + i(\d_xd)\,\sigma_3  +{i\over \lambda} g^{-1}S g=0
\label{EHxiv}
\fin
Projecting eq.(\ref{EHxiv}) on the Pauli matrices $\sig_i$'s
gives differential equations for $v$ and $d$:
\debut
\d_x v &=& -\frac{i}{\la}(S_-+2vS_3 - S_+v^2) \non
\d_x d &=& \frac{1}{2\la}(-2S_3+vS_++{\bar v}S_-)\nonumber
\fin
The first of these equations is a Ricatti equation for $v(x)$.
Expanding in $\la$ the functions $v(x)$ and $d(x)$ as:
\debut
\d_xd(x)\ &=&\ -\frac{s}{\la} + \sum_{n=0}^\infty\ \rho_n(x)\la^n\non
v(x)\ &=&\ \sum_{n=0}^\infty\ v_n(x)\la^n
\qquad;\qquad v_0=\frac{S_3-s}{S_+} \nonumber
\fin
The Ricatti equation (\ref{EHxiv})  becomes:
\debut
2is\, v_{n+1}\ &=&\ -v'_n + iS_+\sum_{m=1}^n\ v_{n+1-m}v_m \non
\rho_n\ &=&\ \frac{1}{2}\({ v_{n+1}S_+ + {\bar v_{n+1}} S_-}\)\label{EHric}
\fin
Note that $v(x)$ is regular at $\la=0$. Equations (\ref{EHric})
recursively determine the functions $v_n(x)$ and $\rho_n(x)$ as
local functions of the dynamical variables $S^i(x)$.
This describes the asymptotic behaviour of $T(\lambda)$ at
$\lambda=0$. The asymptotic series become convergent if we
regularize the model by discretizing the space interval.

Concerning the monodromy matrix $T(\lambda)$, since $g(x)$ is
local and if we assume periodic boundary conditions, we can write
\debut
T(\lambda)= \cos P_0(\lambda)~{\rm Id}
+i \sin P_0(\lambda)~ M(\lambda) \nonumber
\fin
where $M(\la)=M(L;\la)$ with $M(x,\la)=g(x)\sig_3g^{-1}(x)$ and
\debut
P_0(\lambda)=\int_0^L ~dx (\d_x d) \label{EHxxi}
\fin
The trace of the transfer matrix $tr(T(\lambda))$ is:
\debut
tr(T(\lambda ))=2 \cos P_0(\lambda) \label{EHxxii}
\fin
Using the previous result at $\lambda =0$,
we can use $P_0(\lambda)$ as a generating function
for the commuting local conserved quantities:
\debut
I_n = \int_0^L dx \rho_n(x) \nonumber
\fin
The first ones are
\debut
I_0&=&{i\over 4s} \int_0^L dx~\log \( {S_+\over S_-} \)\cdot \d_x S_3
 \non
I_1&=&  -{1\over{16s^3}} \int_0^L dx~ tr(\d_xS \d_xS) \non
I_2&=& {i\over{64 s^5}}\int_0^L dx~ tr(S\[\d_xS,\d_x^2S\])
\nonumber
\fin
$I_0$ and $I_1$ correspond to momentum and energy respectively.

\bigskip
\noindent{\bf 3c- Dressing transformations.}
We now describe the group of dressing transformations for the
Heisenberg model. Our first task is to specify the Lie algebras ${\cal
G}_{\pm}$ entering the factorization problem.
The matrices $r^\pm$ correspond to expanding the matrix $r(\la,\mu)$
either in powers of $(\la/\mu)$ or in powers of $(\mu/\la)$; e.g.:
\begin{eqnarray}
r^+(\lambda ,\mu )&=& \sum_{n=0}^\infty \lambda^n \sigma_i \otimes
\left( {{\sigma_i}\over
\mu^{n+1}}\right) \nonumber \\
r^-(\lambda ,\mu)&=&- \sum_{n=0}^\infty \left( {{\sigma_i} \over
\lambda^{n+1}}\right) \otimes (\mu^n \sigma_i ) \nonumber
\end{eqnarray}
These are projection operators in the loop algebra
$\widetilde{su(2)}$.
For any $X(\lambda)=\sum_i \lambda^i X_i \in
\widetilde{su(2)}$, we define as in eq.(\ref{ECvii}),
\begin{eqnarray}
X_\pm (\lambda)= R_\pm(X(\la))=
\oint {{d\mu}\over{2i\pi}}tr_2\BL r_{12}^\pm (\lambda
,\mu) 1\otimes X(\mu)\BR \label{EZi}
\end{eqnarray}
We have
\begin{eqnarray}
X_+ (\lambda)&=& \sum_{n\ge 0} \lambda^n X_n \label{dt1}  \\
X_- (\lambda)&=&- \sum_{n < 0} \lambda^n X_n \label{dt2}
\end{eqnarray}
Therefore,  ${\cal G}_{\pm}={\rm Im}R_\pm$ are the subalgebras of
$\widetilde{su(2)}$ with  elements of the form eq.(\ref{dt1}) and
eq.(\ref{dt2}) respectively. Notice that any  $X(\lambda) \in
\widetilde{su(2)}$ has a unique decomposition
\begin{eqnarray}
X(\lambda ) = X_+(\lambda) -X_-(\lambda) \nonumber
\end{eqnarray}
This defines our factorization (or Riemann-Hilbert) problem.\\
Remark that the transfer matrix $T(x,\la)$ belongs to the
subgroup $G_-=\exp(\CG_-)$. This is an important particularity
of the Heisenberg model.

 From the factorization problem, one can define the dressing
transformations. Let  $X(\lambda) \in \widetilde{su(2)}$ and set
\begin{eqnarray}
\Theta(\la) =\({T X T^{-1} }\)(\la)=
\Theta_+(\la) -\Theta_-(\la) \nonumber
\end{eqnarray}
A dressing transformation is a gauge transformation with either
$\Theta_+$ or $\Theta_-$
\begin{eqnarray}
\delta_X A_x &=&- [A_x, \Theta_+] - \partial_x \Theta_+ \nonumber
\end{eqnarray}
The gauge transformation with $\Th_-$ gives the same result.\\
If $X \in {\cal G}_-$, then $\delta_X S(x) =0$ since $\Theta
=-\Theta_-$.  For $X\in\CG_+$, one has:
\begin{eqnarray}
\delta_X S(x) &=& - \[{S(x),\Theta_+\Big(x)\vert_{\lambda =0}}\] \non
  &=& -\oint \frac{d\la}{2i\pi\la}\ \BBL S(x), \Th(x,\la)\BBR \label{EZii}
\end{eqnarray}
This is proved by noticing that \cite{Ue83}:
\begin{eqnarray}
\delta_X S &=& -i Res_{\lambda =0}(\delta_X A_x) \nonumber \\
   &=& i\partial_x Res_{\lambda =0}(\Theta_+) +iRes_{\lambda
=0}([A_x,\Theta_+]) \nonumber
\end{eqnarray}
One can give a more explicit description of these transformations.
\proclaim Proposition.
Let $X=i\lambda^n v$, with $v^+ =v $, and denote $\de_X$ by $\de^n_v$.
Then:
\begin{eqnarray}
\delta^n_v S(x)\ =\ i\BBL\ Z^n_v(x)\ ,\ S(x)\ \BBR\label{dt4}
\end{eqnarray}
where the functions $Z^k_v(x)$ can be computed recursively by
\begin{eqnarray}
\partial_x Z^k_v(x) +i \BBL S(x),Z^{k-1}_v(x)\BBR =0\qquad;
\qquad Z_v^0 =v \label{dt3}
\end{eqnarray}

This can be proved as follows.
Let us define the functions $Z^k_v(x)$ by:
\begin{eqnarray}
(TvT^{-1})(x,\la) =\sum_{k=0}^\infty \lambda^{-k} Z^k_v(x) \label{dt2.5}
\end{eqnarray}
The differential equations satisfied by these fucntions are consequences
of those satisfied by the transport matrix $T(x,\la)$. For $X=i\la^n v$,
we can expand $\Th(x,\la)$ as:
\begin{eqnarray}
\Theta(x,\la) =i\sum_{k=0}^\infty \lambda^{n-k} Z^k_v(x) \nonumber
\end{eqnarray}
With these notations, the factorization problem has a simple solution:
\begin{eqnarray}
\Theta_+ =i\sum_{k=0}^n \lambda^{n-k} Z_v^k,~~~~\Theta_- =-i\sum_{k=n+1}^\infty
\lambda^{n-k} Z_v^k  \nonumber
\end{eqnarray}
It is then also a simple exercise to check that the form of $A_x$ is
preserved. In fact
\begin{eqnarray}
\delta_X A_x &=&- [A_x, \Theta_+] - \partial_x \Theta_+ \nonumber \\
 &=& \lambda^{-1} [ S, Z^n_v ] -
 i\sum_{k=0}^{n-1} \left\{ \partial_x Z_v^{k+1}
 +i [S, Z_v^k] \right\} \nonumber \\
 &=& \lambda^{-1}  [ S, Z^n_v ] \nonumber
 \end{eqnarray}
the last sum vanishes by vitue of eq.(\ref{dt3}) and we are left with
$\delta^n_v S(x) = i\[ Z^n_v(x),S(x)\]$. This proves eq.(\ref{dt4}).
One can check similarly that the form of $A_t$ is unchanged, and
its variation is compatible with eq.(\ref{dt4}).
It in particular implies that the equations of motion are invariant.
In other words, the transformations (\ref{dt4}) are symmetries of
the equations of motion. This can also be checked directly.

It is proved in the appendix that this action is Lie-Poisson i.e.
\begin{eqnarray}
\{{^gS}_i(x) , {^gS}_j(y) \}_{G\times M}=\epsilon_{ijk}~{^gS}_k(x)
\delta(x-y)
\nonumber
\end{eqnarray}
where for an infinitesimal transformation ${^gS}=S +\delta_X S$.
Taking $X=i\sum_{n\ge 0} \sum_{i=1}^3 \xi_n^i \lambda^n \sigma_i$, the
Poisson bracket on $G$ reads
\begin{eqnarray}
\{ \xi_n^i , \xi_m^j \}_G=\epsilon^{ijk} \xi^k_{n+m+1} \label{EZiii}
\end{eqnarray}
The shift in the grading (i.e. $n+m+1$ instead of $n+m$) is due
to the choice of the bilinear form on ${\tilde {su(2)}}$
made in eq. (\ref{EZi}).

\bigskip
\noindent{\bf 3d- The non-local charges and the monodromy matrix.}
We already remarked that the equations of motion
have the form of a conservation law
\begin{eqnarray}
\partial_t J_t - \partial_x J_x &=&0 \nonumber
\end{eqnarray}
with $ J_t =S$ and $J_x={i\over 2}[S_x,S]$.
Since the dressing transformations are symmetries of the equations
of motion, by dressing this local current we produce new
currents which form an infinite
multiplet of non-local conserved currents,
\begin{eqnarray}
J_t^{n,v} &=& \delta^n_v S = i[Z_v^n,S ] \nonumber \\
J_x^{n,v} &=&- {\textstyle{1\over 2}}[ \partial_x [Z_v^n,S ],S]
-{\textstyle{1\over 2}} [S_x ,[Z_v^n,S]]
\label{denis}
\end{eqnarray}
for any $n\geq 0$ and $v\in su(2)$.  The charges
are defined by:
\begin{eqnarray}
Q^n_v =\int_0^L J_t^{n,v} (x) dx  = Z_v^{n+1}(L)
\nonumber
\end{eqnarray}
Since the currents are non local, the charges are not conserved. We have
\begin{eqnarray}
{d\over dt}Q_v^n = J_x^{n,v} (L) - J_x^{n,v} (0) \nonumber
\end{eqnarray}
or
\begin{eqnarray}
{d\over dt}Q_v^n ={\textstyle{1\over 2}}[[S_x (L),S(L)],Q_v^{n-1}]
-2 i s^2 [Q_v^{n-2}, S(L)] ~~~~~~~n\ge 2
\nonumber
\end{eqnarray}
Nevertheless, these charges are important because,
as we will see, they are the
generators of the dressing transformations.

We now show that they
are equivalent to the knowledge of the monodromy matrix $T$. Let
\begin{eqnarray}
T(\lambda)=\pmatrix{A(\lambda )& B(\lambda ) \cr C(\lambda ) & D( \lambda ) }
\qquad;\qquad AD-BC=1  \nonumber
\end{eqnarray}
Let us define
\begin{eqnarray}
Q_{ij}(\lambda) ={\textstyle{1\over 2}}
tr( T \sigma_i T^{-1} \sigma_j )(\lambda)
=\delta_{ij} +\sum_{p=0}^\infty \lambda^{-p-1} Q_{ij}^p
\nonumber
\end{eqnarray}
where $ Q_{ij}^p ={\textstyle{1\over 2}}tr(Z_{\sigma_i}^{p+1}\sigma_j )$.
The $Q_{ij}(\lambda)$ are the generating functions of the non-local charges.
Introduce the antisymmetric part of the matrix elements $Q_{ij}(\lambda)$
\begin{eqnarray}
Q_i(\lambda) =\sum_{j,k =0}^3 \epsilon_{ijk}Q_{jk}(\lambda) \nonumber
\end{eqnarray}
The quantities $Q_i(\lambda)$ are quadratic functions of the matrix elements
of $T(\lambda)$. But, one can invert these relations and
express $T(\lambda)$ in terms
of the $Q_i(\lambda)$'s;
\proclaim Proposition.
The relation between the transfer matrix and the
non-local charges $Q_i(\la)$ is:
\begin{eqnarray}
T(\lambda)=
{\textstyle{1\over 2}}X(\lambda)\ Id -{\textstyle{i \over 2}} X^{-1}(\lambda)
\sum_i Q_i(\lambda) \sigma_i
\nonumber
\end{eqnarray}
with
\begin{eqnarray}
X(\lambda)= \sqrt{ 2 +2 \sqrt{1-{1\over 4} \vec{Q}^2(\lambda)} } \nonumber
\end{eqnarray}
\par
Using this result, we can express the matrix $Q_{ij}$ in terms
of the charges $Q_i$. Setting
\begin{eqnarray}
{\cal A}(\lambda)= {1\over 2}\pmatrix{ 0 & Q_3 & - Q_2 \cr
		      -Q_3 & 0 & Q_1 \cr
		      Q_2 & -Q_1 & 0 }
\nonumber
\end{eqnarray}
we find
\begin{eqnarray}
Q(\lambda) =  Id + {\cal A} +2 X^{-2} {\cal A}^2 \nonumber
\end{eqnarray}
So, the charges $Q_i(\lambda)$ contains the same amount of information
as the transfer matrix.

We will show in the next section that the charges
$Q_i(\lambda)$ as well as $T(\lambda)$ generate the dressing transformations,
and that under Poisson bracket the $Q_i(\lambda)$'s
generate a Yangian. It is interesting in this context to examine more
closely the relation between the $Q_i(\lambda)$ and the matrix
elements of $T(\la)$.  We have
\begin{eqnarray}
Q_+(\lambda) &=& Q_1(\lambda) +i Q_2(\lambda) = 2i
X[\vec{Q}^2(\lambda)]\  C(\lambda) \label{nlc1} \\
Q_-(\lambda) &=& Q_1(\lambda) -i Q_2(\lambda) = 2i
X[\vec{Q}^2(\lambda)]\  B(\lambda) \label{nlc2}
\end{eqnarray}
In view of eq.(\ref{nlc1},\ref{nlc2}),
one can speculate that, in this case, the algebraic
Bethe Ansatz \cite{Fa82} is nothing else but the construction of
highest weight representations of $Y(su(2))$, the symmetry group of
our model.

Moreover
\begin{eqnarray}
tr~T(\lambda)=X[\vec{Q}^2(\lambda)] \nonumber
\end{eqnarray}
Therefore, $\vec{Q}^2(\lambda)$ is also a generating
function for commuting quantities. Its relation with the generating
function $P_0(\la)$ is:
\debut
\vec{Q}^2(\la)\ =\ 4\sin^2\BL 2P_0(\la)\BR \nonumber
\fin
However, expanding $\vec{Q}^2(\la)$ around $\la=\infty$ gives
non-local commuting quantities while expanding $P_0(\la)$ around
$\la=0$ gives the local commuting quantities. The links between
them are hidden in the subtilities of the analytic properties
of the monodromy matrix. These analytic properties should be encoded
into the representation theory of the Yangian.

As another application of the relation between $Q_i$ and $T$, we give
a formula for the coproduct of the charges $Q_i$. Recall that
following an idea of \cite{LuPo78}, and as is well known in the
formalism of integrable systems \cite{Fa82}, the classical counterpart of the
coproduct on $T$ is given by
\begin{eqnarray}
\Delta T_{ij}& =& (T' T)_{ij} \nonumber
\end{eqnarray}
where $T$ and $T'$ are the transport matrices on two adjacent
intervals. Then
\begin{eqnarray}
\Delta Q_{ij}= {\textstyle{1\over 2}}
tr (T' T \sigma_i T^{-1} T'^{-1} \sigma_j )
\nonumber
\end{eqnarray}
and using $ T \sigma_i T^{-1} = \sum_l Q_{il} \sigma_l$,
we obtain
\begin{eqnarray}
\Delta Q_{ij} =  \sum_l Q_{il} Q'_{lj} \nonumber
\end{eqnarray}
Inserting the precise form of $Q$ in terms of ${\cal A}$, we find
\begin{eqnarray}
\Delta {\cal A} &=& {\cal A} + {\cal A}' +
{\textstyle{1 \over 2}}[{\cal A},{\cal A}'] +
X^{-2} ({\cal A}^2 {\cal A}'+{\cal A}'{\cal A}^2) \nonumber\\
&& +X'^{-2} ({\cal A}{\cal A}'^2+{\cal A}'^2 {\cal A})+2
X^{-2} X'^{-2} [ {\cal A}^2, {\cal A}'^2 ] \nonumber
\end{eqnarray}
If we expand in $\lambda$,
$Q_i(\lambda ) = \sum_{n=0}^\infty \lambda^{-n-1} Q_i^n$, we get
\proclaim Proposition.
For $n=0$ and $1$, we have:
\begin{eqnarray}
\Delta Q^0_i &=& Q_i^0 + {Q'}_i^0 \nonumber \\
\Delta Q_i^1 &=& Q_i^1 + {Q'}_i^1 -{\textstyle{1\over 4}} \epsilon_{ikl}
Q^0_k {Q'}^0_l \nonumber
\end{eqnarray}
These are exactly the comultiplications in the Yangian,
cf. eq.(\ref{EHi}).
\par

\bigskip
\noindent{\bf 3e- Charges as generators of dressing transformations.}
We now calculate the Poisson bracket of the charges $Q_i$ with the
dynamical variable $S$. We will first show that the non-Abelian
Hamiltonian of the transformations (\ref{EZii}) is the monodromy
matrix and then, we show that the charges $Q_i$ also generate
these transformations. We start form
\begin{eqnarray}
T_1^{-1}(\lambda,x)\{ T_1(\lambda,x), S_2(y)\} ={1\over{2\lambda}}
\theta(x-y)\sum_{i=1}^3 \sigma_i \otimes [ T(\lambda, y)\sigma_i
T^{-1}(\lambda ,y) ,S(y) ] \label{cgdt1}
\end{eqnarray}
This is easily established from the linear system or from the
definition of $T(\lambda,x)$ as a path ordered exponential. With this
formula, we can check immediately that the generator of the dressing
transformations is $T(\lambda)$ itself. Indeed, setting $x=L$, multiplying
eq.(\ref{cgdt1}) by $X$ on the first space and taking the trace, we
get
\begin{eqnarray}
tr_1\BL X_1 T^{-1}_1(\lambda)\{T_1(\lambda),S_2(y)\}\BR={1\over
\lambda}[T(\lambda,y) X T^{-1}(\lambda ,y),S(y)] \nonumber
\end{eqnarray}
Choosing $X=i\lambda^n v$ and using eqs.(\ref{dt2.5},\ref{dt4}), we find
\begin{eqnarray}
\delta_v^n S(y)=\oint {{d\lambda }\over{2i\pi}} tr_1\BL \lambda^n iv_1
T_1^{-1}(\lambda)\{ T_1(\lambda) ,S_2(y) \}\BR \nonumber
\end{eqnarray}
which is exactly what was to be expected: $T(\lambda) $ is the generator of
dressing transformations.\\
To express this formula in terms of the charges $Q_i$, one  can
also calculate the Poisson bracket of $\Theta_v =TvT^{-1}$ and $S$.
We find
\begin{eqnarray}
\{ Q_{ij} , S (y) \} = \epsilon_{ijk}\delta_k^n S(y)
- \sum_{p=0}^{n-1} \epsilon_{ikl} Q_{lj}^{n-p-1} \delta_k^p S(y) \nonumber
\end{eqnarray}
or else
\begin{eqnarray}
\delta^n_{i} S(y)= {\textstyle{1\over 2}} \{ Q^n_i , S(y) \}
 +{\textstyle{1\over 2}}\sum_{p=0}^{n-1}\left[ Q^{n-p-1}-tr(Q^{n-p-1})Id
\right]_{ik} \delta_k^p S(y)
\nonumber
\end{eqnarray}
This equation can be interpreted in two different ways: (i)
this is a system of equations allowing to express recursively $\delta_v^n
S(y)$ in terms of Poisson brackets; or (ii) it allows to express the
dressing $\de_XS$ in a non-linear way in terms of the charges $Q_i$
as in eq. (\ref{EIxv}).
\proclaim Proposition.
For $n=0$ or $1$, we have:
\begin{eqnarray}
\delta^0_{i} S(y)&=&{\textstyle{1\over 2}}  \{ Q^0_i,S(y) \} \non
 \delta_i^1 S(y)&=&{\textstyle{1\over 2}} \{Q_i^1 ,S(y) \}
-{\textstyle{1\over 8}} \epsilon^{ijk} Q_j^0 \{ Q_k^0 ,S(y) \}
\label{EZv}
\end{eqnarray}
These equations are the semi-classical analogue of the adjoint
action of the Yangian, eq.(\ref{EHadj}).
\par

\bigskip
\noindent{\bf 3f- The Poisson algebra of the charges.}
It remains to calculate the Poisson brackets of the charges. For this purpose,
it is enough to calculate the Poisson brackets of $\Theta$.
We find
\begin{eqnarray}
\{ \Theta_v (\lambda ,x)\x \Theta_w ( \mu ,x)\}&=&{1\over 2}
{1\over {\lambda - \mu}}\sum_i
[\sigma^i ,  \Theta_v (\lambda ,x)]\otimes [\sigma^i , \Theta_w ( \mu ,x)]
\nonumber \\
&&-{1\over 2}{1\over {\lambda - \mu}}\sum_i \Theta_{[\sigma^i ,v]}(\lambda ,x)
\otimes \Theta_{[\sigma^i ,w]}(\mu ,x) \nonumber
\end{eqnarray}
Expanding the factors ${1\over{\lambda -\mu}}$ (it does not matter
whether we expand in $ {\mu \over \lambda }$ or in ${\lambda \over \mu }$),
we find
\begin{eqnarray}
\lefteqn{
\{Q^n_{ij}, Q^m_{pq}\} = 2 \Big( \delta_{iq} Q^{n+m}_{pj} -\delta_{jq}
Q^{n+m}_{pi} + \delta_{ip}Q^{n+m}_{jq} - \delta_{jp}Q^{n+m}_{iq}
\Big)}
 \nonumber \\
&&-2\sum_{a=0}^{n-1} \Big( \delta_{jq}Q^{n-1-a}_{ir}Q^{m+a}_{pr}
- Q^{n-1-a}_{iq} Q^{m+a}_{pj}
-\delta_{ip}Q^{n-1-a}_{rj} Q^{m+a}_{rq} + Q^{n-1-a}_{pj}Q^{m+a}_{iq} \Big)
\nonumber
\end{eqnarray}
Multiplying by $\epsilon_{ijk} \epsilon_{pql}$, we get
\begin{eqnarray}
\lefteqn{
\{Q^n_k , Q^m_l \} = 8 {\cal A}^{n+m}_{kl}}
\nonumber \\
&&-4 \sum_{a=0}^{n-1}
\Bigg[ (\Sigma^{n-1-a}{\cal A}^{m+a})_{kl}
- ({\cal A}^{n-1-a}\Sigma^{m+a})_{kl}
+tr(\Sigma^{m+a}) {\cal A}^{n-1-a}_{kl} - tr(\Sigma^{n-1-a})
{\cal A}^{m+a}_{kl} \Bigg] \nonumber
\end{eqnarray}
where $\Sigma$ denotes the symmetric part of $Q$:
$\Sigma (\lambda)= 2X^{-2}{\cal
A}^2 = \sum_{n=0}^\infty \lambda^{-n-1}\Sigma^n $.\\
For $n=0$ or $1$, we get
\begin{eqnarray}
\{ Q^0_k, Q^m_l \}&=&4 \epsilon_{klr} Q_r^m \non
\{Q^1_k , Q^m_l \}&=&8{\cal A}^{m+1}_{kl}
-4 \Big[ tr(\Sigma^m){\cal A}^0_{kl} -
({\cal A}^0\Sigma^m)_{kl} \Big] \nonumber
\end{eqnarray}
The first equation shows that $Q^0_k$ generate an $su(2)$ subalgebra and
that $Q^m_l$ transform in the adjoint representation.
Setting $m=1$ or $2$ gives
\begin{eqnarray}
\{Q^1_k ,Q^1_l \}&=&4 \epsilon_{klr} Q^2_r
+{\textstyle{1\over 2}} (\vec{Q}^0 \cdot
\vec{Q}^0 ) \epsilon_{klr} Q^0_r \non
\{Q^1_k ,Q^2_l \}&=&4 \epsilon_{klr} Q^3_r + (\vec{Q}^0\cdot \vec{Q}^1)
\epsilon_{klr}Q^0_r +{\textstyle{1\over 2}}\epsilon_{krs} Q^0_s Q^1_r Q^0_l
\nonumber
\end{eqnarray}
With the help of these results it is easy to prove the following:
\proclaim Proposition.
The dressing Poisson algebra
is generated by $Q^0_k$ and $Q^1_l$. These charges
satisfy the semi-classical analogues of the
defining relations of the Yangians:
\debut
\{ Q^0_k, Q^0_l \}&=&4 \epsilon_{klr} Q_r^0 \non
\{ Q^0_k, Q^1_l \}&=&4 \epsilon_{klr} Q_r^1 \label{Esemy} \\
\{ Q_\lambda^1 ,\{Q_\mu^1 ,Q_\nu^0 \}\} - \{Q^0_\lambda \{Q_\mu^1,Q^1_\nu\}\}
&=&  a_{\lambda \mu \nu \alpha \beta \gamma}  Q^0_\alpha  Q^0_\beta Q^0_\gamma
\non
\{\{Q^1_\lambda , Q^1_\mu \},\{Q^0_\rho ,Q^1_\sigma \}\} + \{\{ Q^1_\rho ,
Q^1_\sigma \},\{Q^0_\lambda ,Q^1_\mu \}\}
& = &8 ( a_{\lambda \mu \nu \alpha \beta
\gamma}\epsilon_{\rho\sigma\nu}+ a_{\rho \sigma \nu \alpha \beta
\gamma}\epsilon_{\lambda\mu\nu})Q^0_\alpha Q^0_\beta Q^1_\gamma \nonumber
\end{eqnarray}
The last two equations are the semi-classical Serre relations
since in the semi-classical limit we have
\begin{eqnarray}
\{ Q^0_\al , Q^0_\beta, Q^1_\ga \}_{sym}=Q^0_\alpha Q^0_\beta Q^1_\gamma
 \nonumber
 \end{eqnarray}
Above,
$a_{\lambda \mu \nu \alpha \beta \gamma}= {\textstyle{2\over 3}}
 \epsilon_{\lambda \alpha i }\epsilon_{\mu \beta j}
 \epsilon_{\nu \gamma k} \epsilon_{i j k}$.
\par
These relations are proved directly using all the relations we gave.
There is no extra relation between the charges $Q^0_n$ and $Q^1_k$ since
there is no extra relation between the generators $\de^0_n$ and $\de^1_k$
in the $su(2)$ loop algebra.

Finally, notice that the quantization of the algebra of the dressing
transformations is canonical in the variables $Q^0_n$ and $Q^1_k$
(i.e. one goes from the semi-classical relations to the quantum
relations just by transforming the Poisson brackets into
commutators), but it is not canonical in terms the transfer matrix.

\vfill \eject

\section{Appendix}

We prove that the action of the dressing transformations in the
Heisenberg model is Lie-Poisson.
Let $^gS$ be the dressed spin variable. We want to show
\begin{eqnarray}
\{ ^gS^i(x), ~^g S^j(y) \} = \epsilon^{ijk}~^gS^k (x) \delta(x-y) \nonumber
\end{eqnarray}
or in tensor notation
\begin{eqnarray}
\{ ~^gS_1(x), ~^gS_2(y) \} = {\textstyle{i\over{2}}}[C_{12}, ~^gS_2(x)]
\delta(x-y) \nonumber
\end{eqnarray}
where
$C_{12} = \sum_{i=1}^3 \sigma^i \otimes \sigma^i$
is the Casimir element. Infinitesimally, we have
\begin{eqnarray}
 {}^gS(x) = S(x) + \delta_X S(x) \nonumber
\end{eqnarray}
with
\begin{eqnarray}
\delta_X S(x) = -i\partial_x \oint {dz\over{2i\pi}}\Theta_X (z,x)
\label{deltas}
\end{eqnarray}
We have to check
\begin{eqnarray}
\{ \delta_X S_1(x), S_2(y)\}_M &+& \{ S_1(x),\delta_X S_2 (y)\}_M +
\{ \delta_X S_1(x) , \delta_X S_2 (y) \}_G \non
&=&\ \frac{i}{2}\ \BBL C_{12} , \delta_X S_2 (x) \BBR \delta(x-y)
\label{XXX}
\end{eqnarray}
 where $\{~~\}_M$ is the Poisson bracket on phase space, and $\{~~\}_G$ is
 the Poisson bracket on the group of dressing transformations.
Using eq.(\ref{cgdt1}), we get
\begin{eqnarray}
\{ \Theta_1 (x) , S_2 (y) \}_M =- {1\over{2\lambda}}\theta(x-y)
T_1(x)~ [[T_1^{-1}(y) C_{12} T_1(y),X_1],S_2(y)]T_1^{-1}(x) \label{liepoi3}
\end{eqnarray}
Consider first the term proportional to $\delta(x-y)$ in the left hand side
of eq.(\ref{XXX}). It comes from the derivatives in eq.(\ref{deltas})
acting on
$\theta(x-y)$ in the first two terms of this expression. Explicitly:
\begin{eqnarray}
-{\textstyle{i\over 2}}\delta(x-y)\left\{
\oint {{dz}\over{2i\pi z}} T_1(x) [[T_1^{-1}(x) C_{12} T_1(x) ,X_1],S_2(x)]
T_1^{-1}(x)\right. ~~~~~~~~~~~~~~~~~~~\nonumber \\
 ~~~~~~~~~~~~~~~~~~~~~~~~~-\left.
\oint {{dz}\over{2i\pi z}} T_2(x)
[[T_2^{-1}(x) C_{12} T_2(x) ,X_2],S_1(x)] T_2^{-1}(x) \right\} \nonumber
\end{eqnarray}
or else
\begin{eqnarray}
-{\textstyle{i\over 2}}\delta(x-y) \oint {{dz}\over{2i\pi z}} \left\{
[[C_{12}, \Theta_1 (z,x)],S_2(x)] -  [[C_{12}, \Theta_2 (z,x)],S_1(x)] \right\}
\nonumber
\end{eqnarray}
Using the fact that $[C_{12},S_1(x)]=-[C_{12},S_2(x)]$ and
$[C_{12},\Theta_1]=-[C_{12},\Theta_2]$ we can rewrite this as
\begin{eqnarray}
-{\textstyle{i\over 2}}\delta(x-y) \oint {{dz}\over{2i\pi z}} \left\{-
[[C_{12}, \Theta_2 (z,x)],S_2(x)] +[[C_{12},S_2(x)],\Theta_2(z,x)] \right\}
\nonumber
\end{eqnarray}
Finally, using the Jacobi identity, we get
\begin{eqnarray}
-{\textstyle{i\over 2}}\delta(x-y) \oint {{dz}\over{2i\pi z}}[C_{12},[S_2(x),
\Theta_2(z,x)]]= {\textstyle{i\over 2}} \delta(x-y) [ C_{12} ,
\delta_X S_2 (x)] \nonumber
\end{eqnarray}
Next we consider the terms obtained when the derivatives act on the
other factors. Assume $x>y$. These terms read
\begin{eqnarray}
-{\textstyle{i\over 2}}\partial_x \oint
{{dz}\over{2i\pi z}}T_1(x)[[T_1^{-1}(y)C_{12}T_1(y),X_1],S_2(y)]
T_1^{-1}(x)  \label{check}
\end{eqnarray}
We must compare with
\begin{eqnarray}
\lefteqn{
\{ \delta_X S_1(x), \delta_X S_2 (y) \}_G = } \nonumber \\
&&\partial_x \partial_y
\oint {{dz}\over{2i\pi}} \oint {{dz'}\over{2i\pi}}T_1(z,x) T_2(z',y)
\{ X_1(z) ,X_2(z')\}_G T_1^{-1}(z,x) T_2^{-1}(z',y) =\nonumber \\
&&-i \partial_x \oint {{dz}\over{2i\pi}} \oint {{dz'}\over{2i\pi z'}}
\relax[ S_2(y) , T_1(z,x) T_2(z',y)\{ X_1(z) ,X_2(z')\}_G T_1^{-1}(z,x)
T_2^{-1}(z',y) \nonumber
\end{eqnarray}
Suppose
\begin{eqnarray}
\{ X_1(z) ,X_2(z')\}_G={\textstyle{1\over 2}} \left[ {C_{12}\over{z-z'}},
X_1(z) \right]~~~~ |z'|< |z| \label{Poisson}
\end{eqnarray}
Deforming the integration contour $|z'|<|z|$ to $|z'|>|z|$, we get a
contribution of the residue at $z'=z$.
\begin{eqnarray}
-{\textstyle{i\over 2}}\partial_x \oint {{dz}\over{2i\pi z} }
\left\{[ S_2(y), T_1(x) T_1^{-1}(y) C_{12}
T_1(y) X_1 T_1^{-1}(x) ] \right.~~~~~~~~~~~~~~~~~~~~~~~~~~~~\nonumber \\
\left.-[ S_2(y), T_1(x)  X_1T_1^{-1}(y)  C_{12} T_1(y) T_1^{-1}(x) ] \right\}
\nonumber
\end{eqnarray}
matching exactly eq.(\ref{check}). The remaining integral $|z'|>|z|$ gives
no contribution since we must expand
\begin{eqnarray}
{1\over{z-z'}}= - {1\over{z'}}\sum_{n=0}^\infty \left( {z\over{z'}} \right)^n
\nonumber
\end{eqnarray}
and since there is already a factor ${1\over z'}$, we see that the
order of the pole at $z'=0$ is $\ge 2$.

\vfill\eject

\end{document}